\documentclass[twocolumn,prc,showpacs,showkeys,floatfix]{revtex4}

\usepackage{amsmath,amssymb}
\usepackage{bm}
\usepackage[dvips,final]{graphicx}

\def \G{\mathcal{G}}
\def \T{\mathcal{T}}
\def \R{\mathcal{R}}
\def \P{\mathcal{P}}

\def \A{\mathcal{A}}

\def \e{\varepsilon}
\def \w{\omega}
\def \k{\bm{k}}

\begin{document}
\title{Depletion of the nuclear Fermi sea}
\date{\today}

\author{Arnau Rios}
\affiliation{National Superconducting Cyclotron Laboratory and Department of Physics and Astronomy, Michigan State University,
East Lansing, 48824-1321 Michigan, USA}
\email[]{rios@nscl.msu.edu}

\author{Artur Polls}
\affiliation{Departament d'Estructura i Constituents de la Mat\`eria and Institut
 de Ci\`{e}ncies del Cosmos, Universitat de Barcelona, Avda. Diagonal 647, E-08028 Barcelona, Spain}

\author{W. H. Dickhoff}
\affiliation{Department of Physics, Washington University, 
St. Louis, Missouri 63130, USA}

\begin{abstract}
The short-range and tensor components of the bare nucleon-nucleon interaction induce a sizeable 
depletion of low momenta in the ground state of a nuclear many-body system.
The self-consistent Green's function method within the ladder
approximation provides an \textit{ab-initio} description of correlated
nuclear systems that accounts properly for these effects. 
The momentum distribution predicted by this approach is analyzed in detail, with emphasis on the 
depletion of the lowest momentum state. 
The temperature, density, and nucleon asymmetry (isospin) dependence of the depletion of the 
Fermi sea is clarified. 
A connection is established between the momentum distribution and the time-ordered components of 
the self-energy, which allows for an improved interpretation of the results. 
The dependence on the underlying nucleon-nucleon interaction provides quantitative estimates of 
the importance of short-range and tensor correlations in nuclear systems. 
\end{abstract}


\maketitle

\section{Introduction}

Recent experiments at Jefferson Laboratory have clarified several aspects of the role of 
short-range and tensor correlations in determining the properties of nucleons in the nuclear 
medium.
In an $(e,e'p)$ experiment on ${}^{12}$C, an unambiguous signature of the presence of high-momentum
nucleons was identified~\cite{rohe04}.
In the domain of missing energy and momentum probed by the experiment, the amount of 
single-particle (sp) strength identified corresponded reasonably to scaled theoretical predictions
of the self-consistent Green's function (SCGF) calculation for ${}^{16}$O~\cite{muther95}
and correlated basis functions (CBF) calculations for nuclear matter
in the local density 
approximation~\cite{sick94}.
The underlying mechanism for the presence of these high-momentum components is associated with
the strong repulsion that nucleons experience when they are in close proximity, generating a 
suppression of the relative wave function in coordinate space.
In turn, this repulsion is required by the experimental nucleon-nucleon (NN) phase shifts.
An additional mechanism that provides high-momentum components is the action of the NN tensor
force, mediated to a large extent by the exchange of the pion.

In even more demanding exclusive two-nucleon knock-out experiments on ${}^{16}$O to the ground
state of ${}^{14}$C~\cite{Ond98,starink00}, direct evidence for the presence of short-range
proton-proton (pp) correlations was identified, in reasonable agreement with theoretical 
calculations~\cite{giusti98,barbieri04}.
Definitive evidence for the importance of the nuclear tensor force in generating high-momentum
components was presented in Ref.~\cite{subedi08}, where the ratio of knocked-out proton-neutron (pn)
to pp pairs from ${}^{12}$C was found to be around 20.
Theoretical relative momentum distributions exhibit a similar enhancement of pn over pp 
correlations due to the tensor force in the domain of momenta probed in this Jefferson 
Lab experiment~\cite{schiavilla07}.
Possible implications for the physics of neutron stars were
discussed in Ref.~\cite{frankfurt08}.

Particle number conservation requires the high-momentum components to
be accompanied by a corresponding depletion of the nuclear Fermi sea.
A characteristic feature of this depletion in nuclear matter is its essentially
momentum-independent character, except in the immediate vicinity of the Fermi 
surface~\cite{Vond91}.
A survey of calculations~\cite{fantoni84,grange87a,benhar90,baldo90} of the momentum distribution 
of nuclear matter at normal density in Ref.~\cite{Vond91} demonstrated that, for all realistic NN 
interactions then available, different
many-body techniques consistently predicted a depletion of the nuclear Fermi sea of a little over 
15\%. About one-third of this depletion results from tensor correlations.
In addition, the influence of three-body forces on the depletion appears to be rather 
insignificant~\cite{fantoni84}.
An $(e,e'p)$ experiment on ${}^{208}$Pb at NIKHEF in a large domain of missing energy and a 
momentum range corresponding to the mean-field Fermi sea confirmed that a global depletion
between 15 and 20\% of proton orbits below the Fermi energy explains all the measured coincidence 
cross sections~\cite{Bat01,Lap08}.
To put these results in perspective it is useful to note that the depletion of liquid 
${}^3$He for very small momenta is considerably larger and reaches about 
50\%~\cite{mazzanti04}, while the depletion of the electron Fermi sea in closed-shell atoms
is essentially zero~\cite{barbieri07}.

While there is uncertainty about the precise nature of the short-range part of the NN 
interaction, there is now evidence from recent lattice QCD
calculations 
that the features of a strong repulsive short-range core emerge from
first principles, particularly when the pion mass is reduced towards more
realistic values~\cite{ishii07}.
The presence of short-range correlations is thus corroborated by QCD
simulations and strongly suggests that fully microscopic nuclear many-body calculations should 
continue to address their consequences in detail.
Recently developed realistic NN interactions, like  
CDBonn~\cite{cdbonn,cdbonna} or the chiral interactions~\cite{n3lo}
have in general 
softer cores than older potentials~\cite{frick05,dickhoff05}. Even modern
local potentials, such as Argonne v18 (Av18)~\cite{av18}, have a soft
core compared to more traditional interactions. 
By assessing the different results for the Av18 and CDBonn interactions in the present paper, 
we plan to develop a measure of the remaining uncertainty of the role
of short-range and tensor correlations in nuclei.

A particularly well-suited technique for this goal is the SCGF method~\cite{dickhoff04}.
Within this approach, a fully self-consistent treatment of ladder diagrams for the interaction 
between particles that propagate with respect to a correlated (by short-range and tensor effects) 
ground state are accounted for. Recently,
calculations including full off-shell effects have become available at finite 
temperature~\cite{libbyphd,frick03,frickphd,riosphd}, providing thermodynamically consistent 
results that preserve sum rules, like \textit{e.g.} conservation of the number of 
particles~\cite{kadanoff,baym62}.
Earlier applications of this approach at $T=0$ involved a discretized version of the spectral
distribution and have challenged the conventional interpretation of the nuclear-matter 
saturation problem~\cite{dewulf03}.
Finite-temperature effects are expected to smooth out the momentum distribution near the Fermi 
momentum, not unlike the influence of pairing
correlations~\cite{dickhoff05}. Thermal effects hardly 
affect the small and very high-momentum content of the ground state, however.
This expectation is confirmed in general by the present work, although it requires some 
qualification in the case of low density or high temperatures, as we
shall discuss in this work.

The study of the asymmetry dependence of short-range and tensor
correlations is motivated to a large extent
by the (future) study of rare isotopes with large neutron excess.
Experimental studies employing heavy-ion knock-out reactions already suggest that 
the removal probability for the minority species is strongly reduced compared to shell-model 
calculations, whereas the majority species exhibits properties that are essentially mean-field 
like~\cite{gade04}.
Similar tendencies, but smaller in magnitude, have been 
obtained for the spectroscopic factors of protons from a combined
dispersive optical model analysis of a sequence of Ca isotopes~\cite{charity06,charity07}.
Fitting a huge collection of elastic proton scattering data while including results obtained 
from $(e,e'p)$ reactions for quantities below the Fermi energy, the emerging complex optical 
potentials for protons exhibit a striking increase in surface absorption with increasing nucleon 
asymmetry.
This translates into a qualitatively similar tendency in spectroscopic factors as obtained from the zero-momentum 
occupation of protons calculated in the bulk for isospin-polarized nuclear 
matter~\cite{frick05}.
A major purpose of the present work is to clarify the role of various variables, 
such as the temperature, the density or the choice of the NN
interaction, in the determination of the momentum distribution for isospin-polarized systems. In this fashion, we can assess the importance of
the short-range and tensor contributions to the bulk properties of asymmetric nuclei.

Recent applications of the SCGF method to isospin imbalanced matter
have generated predictions for 
the depletion of the proton and the neutron Fermi seas at fixed total density as a function 
of nucleon asymmetry~\cite{frick05}.
Some intriguing results were reported in this work that require a deeper understanding.
First, an increasing difference between the zero-momentum occupation of the neutron and proton
Fermi sea was reported at a temperature corresponding to 10 MeV.
Naively, this difference is expected to be associated with the decreasing
(increasing) importance of the nuclear 
tensor force for neutrons (protons) with increasing asymmetry, since neutrons alone do not 
experience the strong ${}^3S_1$-${}^3D_1$ tensor coupling, while
protons will. There are, however, other physical effects that can influence this
difference and it is therefore useful to investigate whether the 
calculated predictions depend on the choice of the realistic
interaction or on the temperature.
Another feature that requires a better understanding is the opposite
density dependence of the depletion for symmetric 
and pure neutron matter (PNM) reported in~\cite{frick05}.
For symmetric nuclear matter (SNM), a \emph{decrease} in the depletion of the zero-momentum state
with increasing density is obtained, while the opposite result pertains for PNM
in the case of the CDBonn interaction~\cite{cdbonn}.
The result for SNM differs from earlier 
calculations, like those reported 
in Ref.~\cite{baldo90} for a separable version~\cite{Hai84} of the
Paris interaction~\cite{paris80}, which predicted very little density dependence of this 
quantity.
We propose to clarify this issue in the present work by analyzing more carefully the 
ingredients that determine the density dependence of the depletion,
paying particular attention to the importance of thermal effects and
to the choice of the interaction.
To provide a better insight into these dependences, we have relied and
further clarified an approximate relation between
the momentum distribution, above and below the Fermi surface, and the
energy derivatives of the time-ordered components of the self-energy.
These components differ from the retarded quantities that have been
used in finite-temperature calculations. 

In Sec.~\ref{sec:form}, we present the relevant ingredients of the formalism of the 
finite-temperature implementation of the SCGF method.
Results for SNM and PNM are presented in Sec.~\ref{sec:mom1}, where
the role of the temperature, the density, and the choice of realistic interaction
(softer \textit{vs.} harder core) are clarified.
A deeper insight into the density and isospin-polarization dependence of the results is obtained
by employing the relation between the momentum distribution and energy derivatives of the 
different time-ordered components of the nucleon self-energy.
This topic is discussed in Sec.~\ref{sec:self}.
Results for isospin-polarized matter are analyzed in Sec.~\ref{sec:mom2} and further interpreted
with the help of the relations discussed in Sec.~\ref{sec:self}.
Finally, conclusions are drawn in Sec.~\ref{sec:conclu}.

\section{Self-Consistent Green's Functions method at finite temperature}
\label{sec:form}

\subsection{Single-particle propagators}

The results that are presented in the following have been obtained with the SCGF method at finite 
temperature. 
In quantum statistical mechanics, the expectation value of any operator $\hat X$ is given by a 
grand-canonical average,
\begin{align}
\langle \hat X \rangle = \frac{\sum_n e^{-\beta( E_n - \mu N_n) }
  \langle n| \hat X | n \rangle}{\sum_n  e^{-\beta( E_n - \mu N_n)}} \, ,
\label{eq:operator}
\end{align}
where $\beta=1/T$ is the inverse temperature and $\mu$ is the chemical potential of the 
system~\cite{kadanoff,fetter,dickhoff}. 
The many-body eigenstates of the system, $| n \rangle$, diagonalize simultaneously the 
Hamiltonian and the particle number operators,
\begin{align}
\hat H | n \rangle &= E_n | n \rangle \, , \\
\hat N | n \rangle &= N_n | n \rangle \, ,
\end{align}
and their Boltzmann sum (the denominator of Eq.~(\ref{eq:operator})) defines the partition 
function of the system, $\mathcal{Z}$. 
These eigenstates contain all the micro- and macroscopic information of the system, but an 
\textit{ab initio} calculation including all of them is intractable. 
The description of the system can be substantially simplified, without loosing physical 
information, by considering particular kinds of excitations on top of the thermal bath. 
At the sp level, for instance, all the information of a homogeneous quantum many-body system is 
encoded in the time-ordered one-body propagator,
\begin{align}
i \G_\T (k;t-t') = \left \langle \T \left[ \hat a_k(t) \hat a^\dagger_k(t') \right] \right \rangle \, ,
\label{eq:gtime}
\end{align}
given by the thermal average of a product of Heisenberg-picture operators associated with the 
addition and removal of sp excitations with momentum $k$ between times $t$ and $t'$. 
The time-ordering is implemented by the Wick operator, $\T$, which arranges the operators in a 
chronological order and incorporates a $+1$ ($-1$) factor according to the even (odd) nature of 
the corresponding permutation \cite{fetter,dickhoff}. 
The ordering of the time variable determines the analytical structure of $\G_\T$ in the Fourier 
space associated with the time difference, $t-t'$.

The correlation functions,
\begin{align}
i \G^> (k;t-t') &= \left \langle \hat a_k(t) \hat a^\dagger_k(t') \right \rangle \\
i \G^< (k;t-t') &= - \left \langle \hat a^\dagger_k(t')  \hat a_k(t) \right \rangle \, ,
\label{eq:ggt_glt}
\end{align}
are equal to the time-ordered propagator in the corresponding $t>t'$ and $t<t'$ domains. 
Because of the absence of time-ordering, $\G^>$ and $\G^<$ are analytical and well-defined 
functions at all energies. 
The Lehmann representation for $\G^>$,
\begin{align}
\G^> (k,\w) = 2 \pi \sum_{n,m} & \frac{e^{-\beta(E_n - \mu N_n)}}{\mathcal{Z}} \times \nonumber \\
& | \langle m | \hat a^\dagger_k | n \rangle |^2 \, \delta( \w - E_m + E_n ) \, ,
\label{eq:ggt_lehmann}
\end{align}
provides a physical interpretation for this function, as the probability of adding a particle 
with momentum $k$ on top of the thermal bath and ending up in any possible final state, $m$, as 
long as the energy difference between the states coincides with $\w$. 
A similar interpretation exists for $\G^<$ in terms of the removal of a particle from the bath. 
For a system in thermal equilibrium, the correlation functions are connected by detailed balance,
\begin{align}
\G^< (k,\w) &= e^{\beta(\w - \mu)} \G^> (k,\w) \, ,
\label{eq:KMS}
\end{align}
yielding the so-called Kubo-Martin-Schwinger (KMS) relation~\cite{kadanoff}. 
The sum of $\G^>$ and $\G^<$ defines the sp spectral function of the system, $\A(k,\w)$, which 
is individually linked to either of them by the relations,
\begin{align}
\G^< (k,\w) &= f(\w) \A (k,\w) \, , \label{eq:glt_asf} \\
\G^> (k,\w) &= \left[1 - f(\w)\right] \A (k,\w) \, , 
\label{eq:ggt_asf}
\end{align}
with $f(\w) = \left[ 1+e^{\beta(\w - \mu)} \right]^{-1}$ the Fermi-Dirac distribution.

The retarded propagator
\begin{align}
i \G_\R (k;t-t') = \Theta(t-t') \left \langle \hat a_k(t) \hat a^\dagger_k(t') \right \rangle \, ,
\label{eq:gretarded}
\end{align}
is related to the causal propagation of perturbations in the system, as enforced by the 
presence of the Heaviside function in relative time. 
Its spectral decomposition in frequency space is given uniquely in terms of the 
spectral function
\begin{align}
\G_\R (k,\w) = \int_{-\infty}^{\infty} \frac{\textrm{d} \w'}{2 \pi} \frac{ \A(k,\w')}{\w_+-\w'} \, ,
\label{eq:spectral_R}
\end{align}
where the notation $\w_\pm = \w \pm i \eta$, with $\eta$ infinitesimally small, has been 
introduced. 
Decomposing Eq.~(\ref{eq:spectral_R}) in real and imaginary parts,
\begin{align}
\textrm{Im} \G_\R(k,\w) &= -\frac{1}{2} \A(k,\w) \, , \\
\textrm{Re} \G_\R(k,\w) &= - \P \int_{-\infty}^{\infty} \frac{\textrm{d} \w'}{\pi} \frac{ \textrm{Im} \G_\R(k,\w') }{\w-\w'} \, , \label{eq:dispersion_R}
\end{align}
a direct connection between $\textrm{Im} \G_\R$ and the sp spectral function is found. 
Moreover, a dispersion relation ($\P$ denotes a principal part integration) links the real part 
of $\G_\R$ to the energy dependence of its corresponding imaginary part.

Similarly, the spectral decomposition of the time-ordered propagator can be written as the sum of 
two terms 
\begin{align}
  \G_\T(k,\w) &=  \G_\uparrow (k,\w) + \G_\downarrow (k,\w) \, ,
\end{align} 
each of them depending both on the spectral function and an additional phase space factor.
The real and imaginary parts of the up,
\begin{align}
\textrm{Im} \G_\uparrow(k,\w) &= \frac{1}{2}  f(\w) \A(k,\w) \, , \\
\textrm{Re} \G_\uparrow(k,\w) &= \P \int_{-\infty}^{\infty} \frac{\textrm{d} \w'}{\pi} \frac{ \textrm{Im} \G_\uparrow(k,\w') }{\w-\w'} \, ,  \label{eq:dispersion_up}
\end{align}
and the down components,
\begin{align}
\textrm{Im} \G_\downarrow(k,\w) &= -\frac{1}{2} \left[ 1 - f(\w) \right] \A(k,\w)  \, , \\
\textrm{Re} \G_\downarrow(k,\w) &= - \P \int_{-\infty}^{\infty} \frac{\textrm{d} \w'}{\pi} \frac{  \textrm{Im} \G_\downarrow(k,\w') }{\w-\w'} \, ,  
\label{eq:dispersion_down}
\end{align}
are linked by independent dispersion relations. 
It is then easy to show that the retarded propagator is related to the
time-ordered components via the expressions: 
\begin{align}
\textrm{Im} \G_\uparrow(k,\w) &= - f(\w)   \textrm{Im} \G_\R(k,\w)  \, , \label{eq:up_r}\\
\textrm{Im} \G_\downarrow(k,\w) &= \left[ 1 - f(\w) \right] \textrm{Im} \G_\R(k,\w)  \, . \label{eq:down_r}
\end{align}

In the zero-temperature limit, the phase space factors in the dispersion integrals guarantee 
that the integration domains for the up and down components are disconnected. 
The integration in $\G_\uparrow$ goes up to the Fermi surface, $\w=\mu$, while the 
dispersion integral for $\G_\downarrow$ starts at this point. 
Since the poles of the up (down) component are in the upper (lower) half of the complex energy 
plane, one can associate the first (second) term to a ``hole'' (``particle'') propagator. 
In contrast, the dispersion integral associated with $\G_\R$ receives simultaneous contributions 
from above and below the Fermi surface. 
As we shall see in the following, this generates rather different energy dependences for the real 
parts of these three quantities. 
Let us also note that the knowledge of the spectral function is enough to compute the real and 
imaginary parts of all the different sp propagators~\cite{mahan}. 

The definition of the time-ordered propagator, Eq.~(\ref{eq:gtime}), can be generalized to purely 
imaginary times with the proper treatment of the imaginary time-ordering~\cite{kadanoff}. 
The following KMS relation arises
\begin{align}
\G_\T(k,\tau) = -e^{\beta \mu} \G_\T(k,\tau- i \beta) ,
\end{align}
which demonstrates that $\G_\T$ is a quasi-periodic function along the imaginary time axis. 
This suggests the following discrete Fourier representation
\begin{align}
\G_\T(k,\tau) = \frac{1}{-i \beta} \sum_\nu e^{-i z_\nu \tau} 
\G(k,z_\nu) \, ,
\end{align}
where $z_\nu=\frac{(2 \nu + 1) \pi}{-i \beta} + \mu$ correspond to fermionic Matsubara 
frequencies \cite{fetter,dickhoff,mahan}. 
An analytical continuation of the coefficients in the Fourier transform, $\G(k,z_\nu)$, to 
continuous complex values of energy can be uniquely defined~\cite{baym61}. 
The spectral decomposition of this function
\begin{align}
\G (k,z) = \int_{-\infty}^{\infty} \frac{\textrm{d} \w'}{2 \pi} \frac{ \A(k,\w')}{z-\w'} \, ,
\label{eq:spectral_G}
\end{align}
is again uniquely given by the spectral function. 
Note that the same function, $\G$, contains information on both the time-ordered propagator (when 
computed at $z=z_\nu$) and the retarded propagator (at $z=\w_+$). 
In general, the sums over Matsubara frequencies can be transformed into energy integrals for the 
retarded components using complex analysis techniques. 

In momentum-frequency space, the Dyson equation for the sp propagator, 
\begin{align}
\left[ z - \frac{k^2}{2m} - \Sigma(k,z) \right] \G(k,z) = 1 \, ,
\label{eq:dyson}
\end{align}
reduces to an algebraic equation in terms of the self-energy, $\Sigma(k,z)$. 
As a consequence of this equation, the self-energy inherits the analytical properties of the 
propagator. 
In particular, it can be decomposed in equivalent retarded, time-ordered or ``less/greater than'' 
contributions~\cite{mahan}. 
Taking the $z \to \w_+$ limit in this last expression leads to a Dyson equation relating the 
retarded Green's functions and the retarded self-energy. 
Since these retarded components form a closed set of equations by themselves, they have been 
extensively discussed, particularly in the nuclear context at 
finite temperature~\cite{schnell96,bozek99,frick03}. 
If one needs to distinguish between particle and hole contributions,
however, time-ordered components become essential~\cite{schnell96}. 
The decomposition in terms of up and down self-energies also allows for a more natural 
connection with zero-temperature SCGF calculations~\cite{ramos89,dewulf02}.

\subsection{Ladder approximation}

The retarded self-energy fulfills the dispersion relation 
\begin{align}
\textrm{Re} \Sigma_\R(k,\w) = \Sigma_{HF}(k) -  \P \int_{-\infty}^{\infty} \frac{\textrm{d} \w'}{\pi} \frac{  \textrm{Im} \Sigma_\R(k,\w') }{\w-\w'} \, . 
\end{align}
The first term in this expression corresponds to an energy-independent Hartree-Fock contribution
\begin{align}
\Sigma_{HF}(k) =  \sum_{\k'}
\left\langle \k \k' | V | \k \k' \right\rangle_A n(k') \, ,
\label{eq:reshf}
\end{align} 
where we have introduced the NN potential (properly antisymmetrized) and the momentum 
distribution, $n(k)$. 
The latter includes correlation effects and can be computed from the sp spectral function 
\begin{align}
n(k) =  \nu \int_{-\infty}^\infty \frac{\textrm{d} \w}{2 \pi} \A(k,\w) f(\w) \, ,
\label{eq:momdis}
\end{align} 
where $\nu=4$ ($\nu=2$) accounts for the degeneracy of nuclear (neutron) matter. 
The energy-dependent, dispersive contribution to the self-energy describes many-body processes 
that go beyond the basic mean-field approximation. 
In the ladder approximation, one includes an infinite series of collisions between a particle and 
a series of particles and holes in the medium, thus accounting for the two-body scattering 
problem in the medium~\cite{fetter,dickhoff}. 
The imaginary part of the retarded self-energy in this approximation reads
\begin{align}
\textrm{Im} \Sigma_\R(k,\w)= \sum_{k'} & \int_{-\infty}^{\infty} \frac{\textrm{d} \w'}{2 \pi}
 \left[ f(\w') + b(\w+\w') \right] \times  \nonumber \\
& \A(k',\w') \left\langle \k \k' | \textrm{Im} T (\w+\w'_+) | \k \k' \right\rangle , 
  \label{eq:imself}
\end{align}
and is given in terms of the in-medium interaction, the sp spectral function and phase space 
factors, including a Bose-Einstein distribution, 
$b(\Omega) = \left[ e^{-\beta(\Omega - 2\mu)} - 1 \right]^{-1}$. 
The $T$-matrix in the medium fulfills a Lippmann-Schwinger-like equation
\begin{align}
\left\langle \k_1 \k_2 | T (\Omega_+) | \k_3 \k_4 \right\rangle_A & =
\left\langle  \k_1 \k_2 | V | \k_3 \k_4 \right\rangle_A \nonumber \\
 + \sum_{\k_5, \k_6} &  \left\langle \k_1 \k_2 | V | \k_5 \k_6 \right\rangle_A\G^0_{II}(k_5,k_6; \Omega_+) \times \nonumber \\
& \left\langle \k_5 \k_6 | T (\Omega_+) | \k_3 \k_4 \right\rangle_A \, ,
  \label{eq:lippschw}
\end{align}
where the retarded $\G^0_{II}$ is obtained from the product of two spectral functions and a phase 
space factor describing the intermediate propagation of particle-particle and hole-hole pairs
\begin{align}
 \G^0_{II}(k,k'; \Omega_+) =  
\int_{-\infty}^{\infty} \frac{\textrm{d} \omega}{2 \pi} \frac{\textrm{d} \omega'}{2 \pi} & \A(k,\w) \A(k',\w') \times \nonumber \\
& \frac{1 - f(\w) - f(\w')}{\Omega_+ - \w -\w'} \, ,
  \label{eq:g20}
\end{align}
properly accounting for Pauli blocking effects at finite temperature~\cite{dickhoff}. 
Once the real and imaginary parts of the self-energy are computed, one can feed back this 
information into the spectral function via the Dyson equation
\begin{align}
\A(k,\w) = \frac{-2 \textrm{Im} \Sigma_\R(k,\w)}{ \left[\w - \frac{k^2}{2m} - \textrm{Re} \Sigma_\R(k,\w) \right] ^2 + \left[ \textrm{Im} \Sigma_\R(k,\w) \right]^2 } \, ,
\label{eq:sf}
\end{align} 
which can in turn be used to compute a new $\G_{II}^0$. 
By iterating this procedure until convergence, the SCGF result within the ladder approximation is 
obtained. 
The importance of self-consistency arises from the fact that sum rules are 
preserved~\cite{polls94,frick04b} and also that thermodynamical consistency is 
fulfilled~\cite{bozek01,rios06}. 
In addition, self-consistency represents a democratic treatment of all the particles considered 
in the problem: the one for which the self-energy is calculated, as well as the ones it interacts 
with.
In all the previous expressions, a grand-canonical ensemble has been assumed, so that there are 
two external, fixed variables, the temperature and the chemical potential. 
The latter is not particularly well suited for in-medium studies, so it is customary to 
supplement the SCGF equations with an equation that fixes the total density of the system
\begin{align}
\rho = \nu \sum_k n(k,\mu) \, ,
\label{eq:norm}
\end{align}
where we have highlighted the dependence of the momentum distribution on the chemical potential 
via the Fermi-Dirac factor of Eq.~(\ref{eq:momdis}). 

The numerical solution of the SCGF equations is a demanding task, due to two major difficulties. 
The first issue is related to the possibility of a pairing solution below a certain critical 
temperature, $T_c$, since nucleons tend to form Cooper pairs in certain 
density regimes ~\cite{thouless60,alm96}. 
Although the SCGF formalism is capable of dealing with the description of the superfluid phase,
we will only discuss results for the normal phase, since the effects of pairing are 
concentrated in a narrow energy region (related to the gap) around the chemical 
potential~\cite{dickhoff05}.
Any conclusion drawn for energies and momenta away from this region should therefore also be 
valid in the pairing regime and, most interestingly, in the zero-temperature limit. 
The second numerical difficulty implementing the SCGF method is related to the requirement to 
solve the coupled nonlinear equations in a wide range of momenta and energies.
Since no quasi-particle assumption is made, complete off-shell propagation effects are included.
In this case, 
different regions in the energy and momentum domains have to be
sampled simultaneously, particularly the quasi-particle peak and 
the high-momentum and energy components of the spectral functions. 
The smoothing of structures associated with finite temperature is helpful in this direction, 
but the problem is still formidable. 
It has taken some time before complete, satisfactory numerical results have been obtained for 
realistic NN interactions. 
We refer the reader to Refs.~\cite{frick03,libbyphd,frickphd,riosphd} 
for further details on how the SCGF is implemented in practice. 

Once a self-consistent solution has been obtained, one has access to the spectral functions of a 
nuclear system at a given density and temperature. 
In addition to the sp properties, the spectral function also determines to a large extent the 
macroscopic properties of the system. 
The total energy, for instance, can be computed from the Galitskii-Migdal-Koltun sum 
rule~\cite{migdal58,koltun74}.
An approximation to the entropy beyond the quasi-particle approximation can also be obtained via 
the Luttinger-Ward formalism~\cite{luttinger60b,rios06}. 
All the remaining thermodynamical properties of the system are
therefore accessible and include properly the effect of short-range and tensor correlations.

So far, we have discussed uniform nuclear systems with a single species: the nucleon, in the case 
of SNM, and the neutron, in the case of PNM. 
For a fixed total density, one can switch from one system to the other by modifying the relative 
concentration of neutrons and protons. The asymmetry parameter
\begin{align}
\alpha = \frac{\rho_n - \rho_p}{\rho_n + \rho_p} \, ,
\end{align}
is a measure of the isospin imbalance. 
The SCGF method within the ladder approximation can be generalized to the case of asymmetric 
nuclear matter and partially isospin-polarized systems can thus be 
analyzed~\cite{bozek04,frick05,rios06b}. 
Since the approach accounts for both the short-range and tensor correlations associated with the 
underlying NN interaction, it can be used to generate quantitative predictions for the importance 
of different types of correlations in isospin asymmetric systems. 
In this sense, it is a unique theoretical tool. 
Other theoretical formalisms either cannot be generalized to isospin asymmetric systems 
(as is the case of the variational approach~\cite{mukherjee07}) or lack adequate consistency 
constraints (as in the case of Brueckner-Hartree-Fock theory~\cite{lejeune86}). 

In the following, we shall employ different microscopic NN interactions to quantify the 
uncertainty related to their short-range and tensor properties. 
Most of the calculations have been performed with the CDBonn~\cite{cdbonn} and the 
Av18~\cite{av18} potentials. 
These phase-shift equivalent forces are representative of two subsets of realistic NN potentials. 
The first is a boson-exchange potential, with a soft short-range core and a somewhat small tensor 
component, as indicated by a relatively low $D$-state probability for the deuteron. 
The latter is a local potential, with a harder short-range core and a more significant tensor 
coupling. 
Although these differences do not affect the two-body scattering observables (which are identical 
for any phase-shift equivalent interaction), they do influence the in-medium properties due to 
their different off-shell behaviors~\cite{muther00}. 
To assess the importance of the different components of the nuclear force, we shall also discuss 
results obtained from the systematically simplified family of NN interactions obtained by the 
Argonne group~\cite{Wiringa02}. 
Each of these forces (from v18 to v4') has a successively simpler operatorial structure, 
with their parameters refitted to describe the two-body system at their corresponding level of 
simplicity. 
Finally, the Reid93 potential, fitted to the Nijmegen partial wave analysis, has also been used 
as a benchmark \cite{stoks94}. 
The momentum distributions obtained with these modern interactions will also be compared to 
results for older realistic interactions~\cite{Vond91}, that have somewhat stronger repulsive cores.
In all cases, our present results have been obtained with partial waves up to $J=8$ in the 
Hartree-Fock and $J=4$ in the dispersive contributions to the self-energy. 

\section{Momentum distribution of one-component nuclear systems\label{sec:mom1}}

\begin{figure}[t]
  \begin{center}
    \includegraphics*[width=\linewidth]{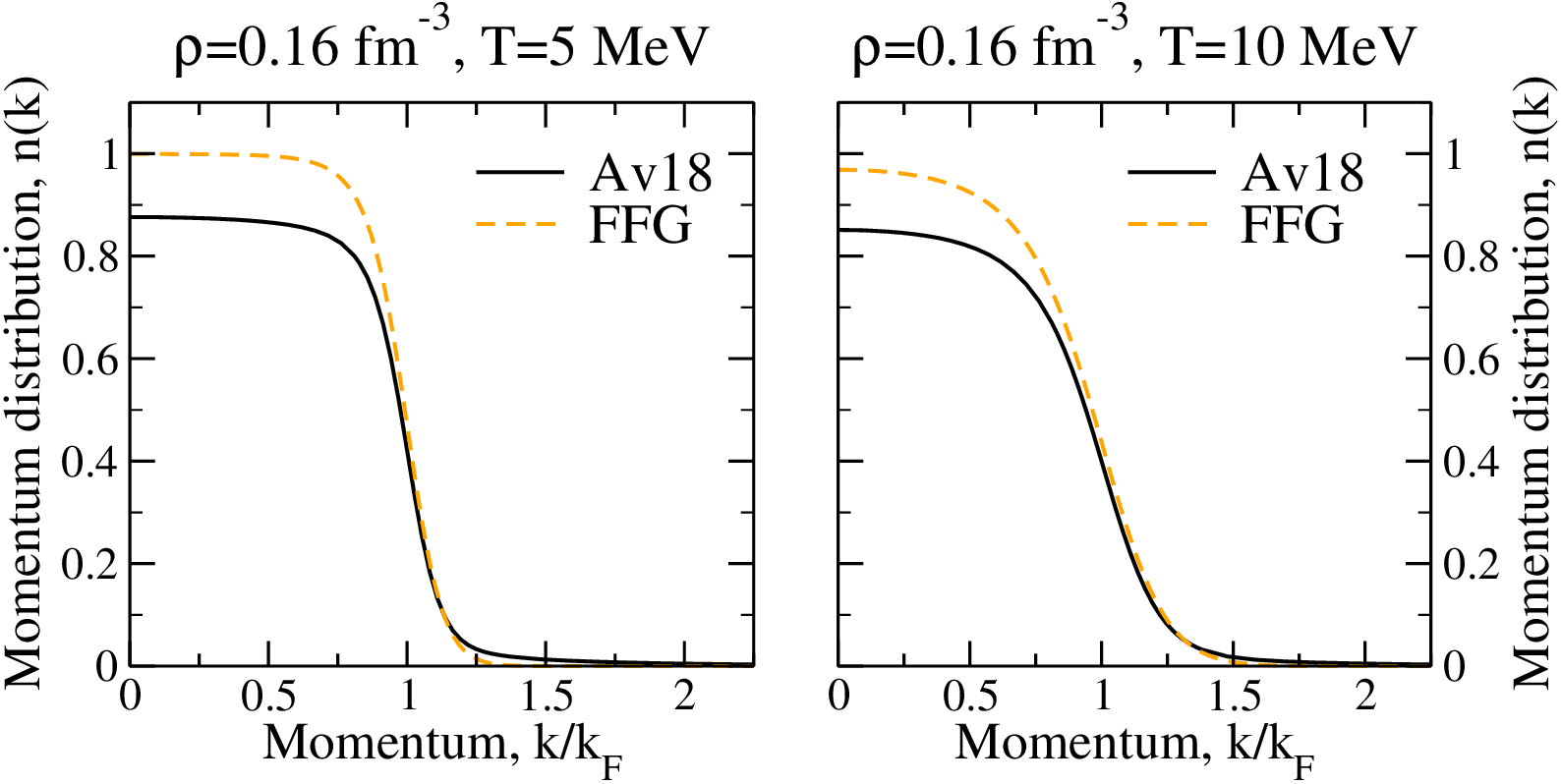}  
    \caption{(Color online) Correlated momentum distribution obtained with the Av18 interaction at $\rho=0.16$ fm$^{-3}$ and two temperatures: $T=5$ MeV (left panel) and $T=10$ MeV (right panel). The dashed line corresponds to the momentum distribution of the Free Fermi Gas under the same conditions.}
    \label{fig:momsym}
  \end{center}
\end{figure}

We start the discussion of the SCGF results by highlighting the effect of correlations on the 
momentum distribution in SNM.
Interaction-induced correlations have a distinctive signature in the momentum distribution, 
removing strength at momenta below the Fermi surface and shifting it to high momenta.
In addition to the correlations induced by the NN force, however, the SCGF momentum distribution 
is sensitive to the temperature of the system. 
As a matter of fact, thermal correlations produce a similar signature on the momentum 
distribution: low momenta are depopulated, while high momenta are thermally occupied. 
The amount of low-momentum depletion and high-momentum population,
however, is significantly different for both types of correlations. 
One can distinguish between these two types of effects by comparing
the fully correlated SCGF prediction 
with the Free Fermi Gas (FFG) results, which is only sensitive to thermal 
correlations~\cite{rios09}. 
To this end, in Fig.~\ref{fig:momsym} we show $n(k)$ of SNM for 
Av18 and the FFG for $\rho=0.16$ fm$^{-3}$ at $T=5$ MeV (left panel) and 10 MeV (right panel). 
At the lower temperature, the Fermi-Dirac distribution of the FFG
deviates very little from a step function, $\Theta (k_F- k)$, which describes 
the $n(k)$ of the FFG at zero temperature. 
At this temperature and density, one can conclude that the thermal effects are small in the deep 
interior of the Fermi sea.
Thermal effects, however, modulate $n(k)$ close to the Fermi surface, which is no longer 
discontinuous at $k_F$. Temperature is also responsible for the fact
that, even in this very degenerate regime, the contribution of the
states below $k_F$ to the density sum-rule is only $86 \%$. The
remaining $14 \%$ corresponds to thermally populated states above $k_F$. 
For the correlated case, $n(k)$ is rather flat below $k_F$ and 
it presents a sizeable depletion (that we will later on characterize
by $n(0)$). With the comparison with the FFG, one can conclude that
dynamical NN correlations are mainly responsible for the depletion. 
Due to these effects, the contribution to the density sum-rule of the
states below $k_F$ is reduced to $75 \%$, \emph{i.e.} $10 \%$ of the
strength is shifted to higher momenta due to NN correlations.
Consequently, momentum components of the wave function above $k_F$ are
generated. The high-momentum tail in $n(k)$ at this density and
temperature provides $52 \%$ of the total kinetic energy per
particle, to be compared to the much smaller $25 \%$ in the FFG.

At a larger temperature (right panel), thermal effects do not only
modify the FFG $n(k)$ around the Fermi surface, but they 
also produce a depletion deep inside the Fermi sea. The lowest
momentum state is depopulated by a few percent, $n(0) =0.97$. 
The shape of $n(k)$ differs significantly from the step function,
documenting the loss of degeneracy at this temperature and density,
even though it is still far from the classical Boltzmann momentum distribution.
The fraction of particles that occupy states below $k_F$ in the FFG decreases to $73 \%$, while 
they contribute only the $53 \%$ of the kinetic energy per particle.
At this temperature, the momentum distribution of the correlated
system deep inside the Fermi sea exhibits a depletion, $1-n(0)=15 \%$, 
which is approximately the sum of the depletion associated 
with the dynamic NN correlations ($\sim 13 \%$)  
plus the one coming from the thermal distribution of the FFG ($3 \%$).
The high-momentum tail is however substantially more important in the
correlated case and the fraction of particles in states above the Fermi momentum increases
to $44 \%$, while the contribution to the kinetic energy amounts to $56 \%$.

\begin{figure}[t]
  \begin{center}
    \includegraphics*[width=\linewidth]{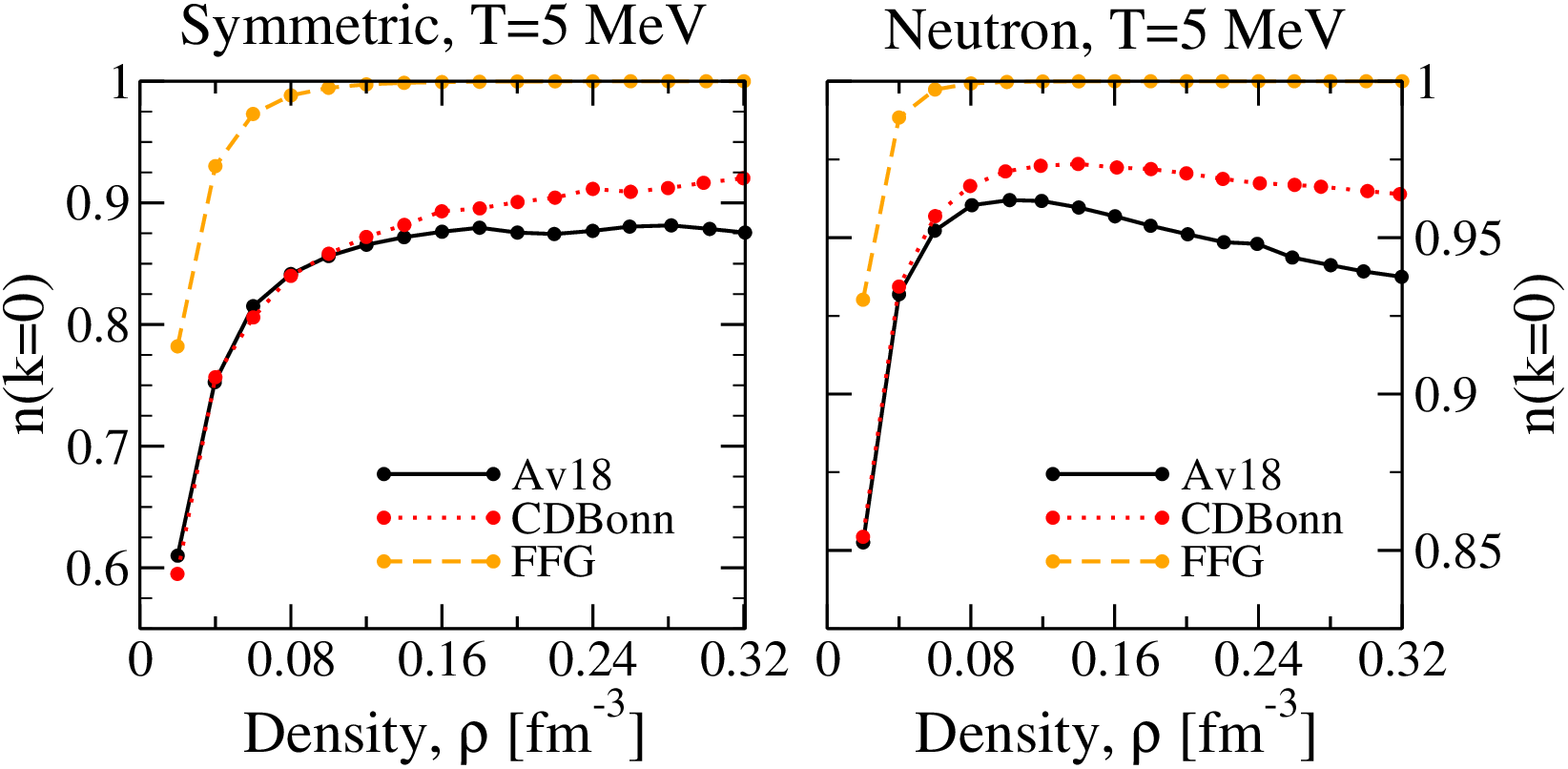}  
    \caption{(Color online) Density dependence of the occupation of the lowest momentum state at $T=5$ MeV for Av18 (solid lines), CDBonn (dotted lines) and the FFG (dashed lines). The left (right) panel corresponds to nuclear (neutron) matter results. Note the different vertical scale of the two panels.}
    \label{fig:depsymneu}
  \end{center}
\end{figure}

The depletion of the momentum distribution below the Fermi surface can be taken as a measure of 
the importance of both thermal and dynamical correlations in many-body systems. 
In the degenerate regime, where thermal effects are unimportant, the
amount of strength removed at low momenta is closely related to the structure of the underlying 
NN force and is particularly sensitive to the short-range core as well as to the tensor 
components. In the correlated case, the low-momentum region of the
momentum distribution is rather flat and almost independent of
momentum as long as the temperature is relatively low and the density
is large enough. 
For high temperatures or low densities, $n(k)$ is softened by thermal correlations. 
One can therefore focus on the $k=0$ state to discuss generic
depletion effects in both regimes. 
Figure~\ref{fig:depsymneu} displays the density dependence of the occupation of this state for 
SNM (left panel) and PNM (right panel). 
Results for the Av18 (CDBonn) interaction are shown in solid (dotted) lines, while dashed lines 
represent the FFG depletion under the same conditions. 

The non-interacting depletion is extremely helpful in understanding
the influence of thermal effects on the low-momentum components. 
At high densities, degeneracy dominates over temperature, and $n(0)=1$ as expected from the zero 
temperature FFG momentum distribution. 
This degenerate regime determines the region where interaction-induced correlations represent 
the major contribution to the depletion in the correlated case. 
Consequently, the results obtained in this regime can be taken as
faithful representatives of zero-temperature calculations. 
At lower densities, however, the FFG drops rapidly as density
decreases, indicating that thermal effects dominate the low-momentum components of $n(k)$. 
The decrease observed in $n(0)$ for the correlated case should
therefore be associated with temperature effects rather than with NN correlations. 
In this region, the correlated $n(0)$ yields interaction-independent results, as expected from 
the virial expansion~\cite{schwenk06} and illustrated in the figure. 
Note also that the onset density for the thermally dominated region is larger for symmetric 
matter ($\rho \lesssim 0.16$ fm$^{-3}$) than for neutron matter ($\rho
\lesssim 0.08$ fm$^{-3}$). This corresponds to the notion that, for a given density, neutron matter is more ``degenerate'', 
\emph{i.e.} has a larger Fermi momentum for the same value of $\rho$. 

If we take the deviation of $n(0)$ from $1$ as a measure of correlations, we can say that neutron 
matter is ``less correlated'', in general, than nuclear matter (note the different vertical scale 
of the two panels of Fig.~\ref{fig:depsymneu}). 
Moreover, the behavior of $n(0)$ as a function of density in the region where thermal effects are 
unimportant is very different for PNM and SNM.
In neutron matter, as the density increases, the population of low momenta drops. 
This result agrees with the intuitive idea that, as particles are
closer together on average, 
the effect of the short-range core increases and low-momentum strength
is shifted to high momenta. Also in accordance with this picture, the depletion is more important 
for an interaction with a harder 
core (Av18) than for a softer force (CDBonn)~\cite{dickhoff05}. 
In stark contrast, for SNM the opposite density dependence is observed: the depletion is 
constant or \emph{decreases} as the density increases. 
For Av18, the depletion saturates at a constant value of $n(0) \sim
0.87$ as density increases, while for CDBonn a clear increase of
$n(0)$ with density is observed. 

What causes such a different behavior in the density dependence of $n(0)$? 
The major difference between neutron and nuclear matter lies in the role of the strong 
$^3S_1-^3D_1$ tensor component, which is only active in the latter. 
It appears that the effect of this component is twofold. On the one
hand, it increases, in general, the depletion as compared to the
effect of short-range correlations by themselves, as seen by comparing
SNM and PNM results. On the other hand, tensor components seemingly
modify the density dependence of $n(0)$. 
This nontrivial result requires a deeper understanding for which the Green's functions formalism 
offers a unique perspective by connecting directly the momentum distribution and the in-medium 
self-energy, as will be discussed in Sec.~\ref{sec:self}. 
Suffice it to say for now that the effect of the tensor force is not only limited to the depletion 
low-momentum components but is also responsible in higher order for binding effects, as 
illustrated by the deuteron.

\begin{figure}[t!]
  \begin{center}
    \includegraphics*[width=\linewidth]{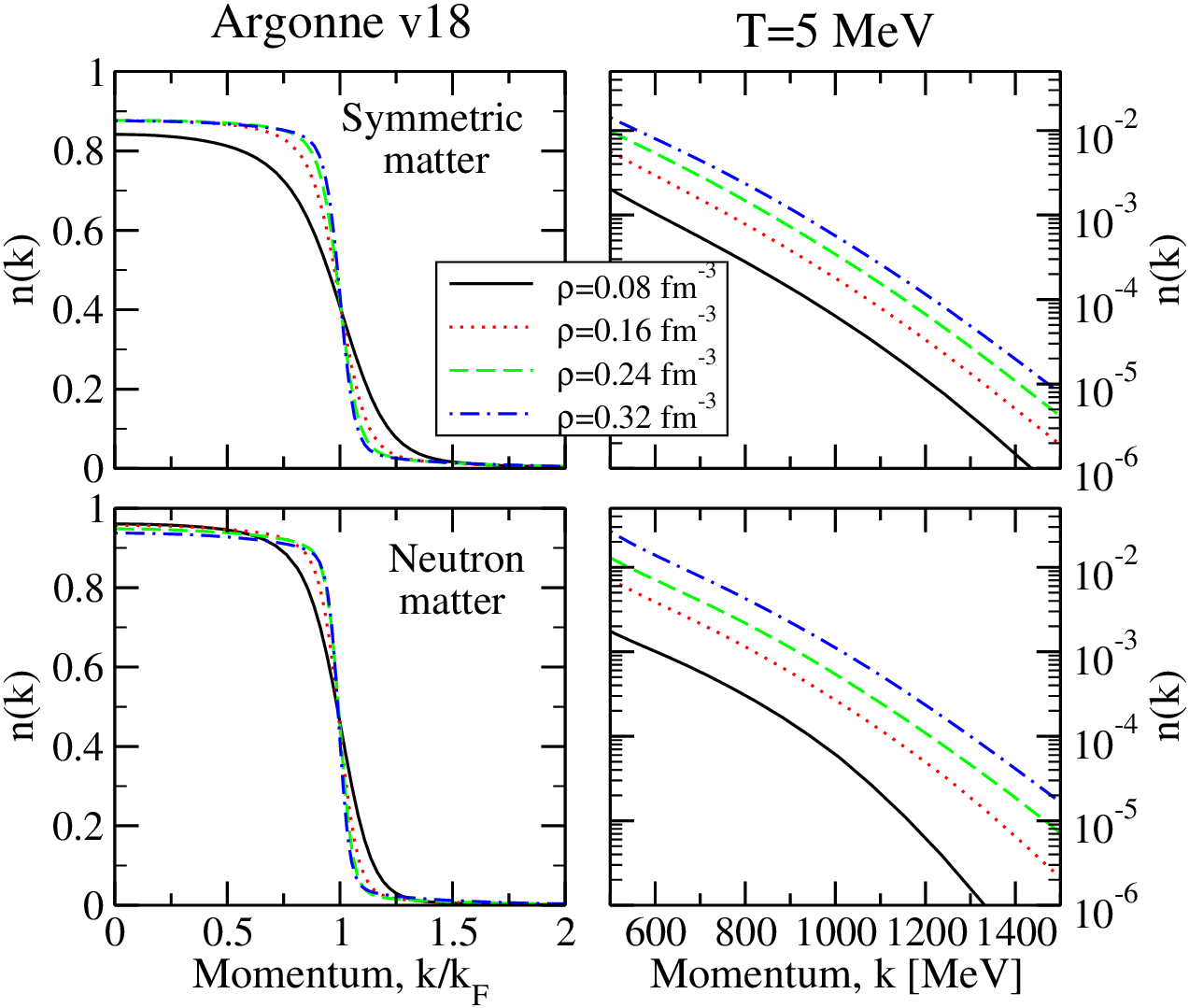}  
    \caption{(Color online) Momentum distribution of symmetric (upper panels) and neutron (lower panels) matter obtained with the Av18 interaction at $T=5$ MeV for different densities. The right panels focus on the high-momentum region.}
    \label{fig:mom_av18_symneu}
  \end{center}
\end{figure}

The normalization of the momentum distribution to the total density, Eq.~(\ref{eq:norm}), forces 
the missing strength at low momentum to be shifted to high-momentum states. 
The shift in strength may depend on the physical properties of the system (temperature, 
density, isospin content). 
Figure~\ref{fig:mom_av18_symneu} shows the momentum distribution of symmetric (upper panels) 
and neutron (lower) matter for Av18. 
In all cases the temperature is $5$ MeV and four different densities are displayed. 
In the left panels the momentum distribution is displayed as a function of the ratio $k/k_F$, 
where $k_F=(6 \pi^2 / \nu \rho)^{1/3}$ is the Fermi momentum of the system. 
This allows for a clear inspection of the density dependence of $n(k)$
and can be instructive in 
understanding the scaling of the distribution as a function of density. 
At a constant temperature, an increase in density leads to a more degenerate system. 
Consequently, the high density results have a more
zero-temperature-like structure, with an almost 
constant depletion below $k_F$ and a substantial jump in $n(k)$ at $k=k_F$. 
In contrast, at low densities, the momentum distribution is modulated
by temperature at all momenta. 
Comparing the four different densities, one can observe that one major
density effect is the redistribution of strength in the region close to the Fermi surface.
Note that on the present scale the effects on the depletion shown in Fig.~\ref{fig:depsymneu} 
are hard to see. 
In general, though, Av18 leads to a saturation of the depletion in symmetric matter and to a 
decrease of the population of low momenta for high-density neutron matter. 

The high-momentum components are shown in the right panels of 
Fig.~\ref{fig:mom_av18_symneu} on a logarithmic scale for momenta 
ranging between $500$ and $1500$ MeV. 
In both symmetric and neutron matter, the high density results lead to a larger population of 
high-momentum states. 
This is in contrast to the low-momentum states, which exhibit a very
different density dependence for the two cases.
All in all, this suggests that the mechanism that produces high-momentum components only scales 
with density and is rather momentum independent. 
There have been suggestions in the literature that high-momentum components scale with $k_F^5$ as 
a function of density. 
Such results have been obtained for a separable version of the Paris interaction~\cite{baldo90} 
and a central (no tensor effects) version of the Reid soft-core interaction~\cite{ramos91}. 
These interactions have a stronger core, which may explain the different behavior.
In addition, in the limit of extremely dilute matter, the high-momentum components scale with 
$k_F^6$ for the hard-sphere Fermi gas~\cite{sartor80}. 
Our calculations, based on more recent, softer realistic interactions, scale more like 
$k_F^3$~\cite{frickphd}.

\section{Relation of $n(k)$ to the time-ordered self-energy \label{sec:self}}

The Green's functions method is unique in that it can provide an interpretation of the previous 
results in terms of the sp spectral function. 
The interplay between thermal and density effects in $n(k)$ can be qualitatively understood in 
terms of the evolution in temperature and density of both the spectral function and the 
phase-space distribution (see Eq.~(\ref{eq:momdis}))~\cite{riosphd}. 
Unfortunately, it is not straightforward to go beyond this sort of reasoning and find a simple 
connection to the in-medium interaction. 
A different perspective and additional insight can be achieved by considering an approximate 
relation between the momentum distribution and the components of the time-ordered self-energy.
The analysis will be performed for $T=0$ to make contact with earlier work on this 
subject~\cite{Huf72,ramos91,baldo91,baldo92,mahaux92}.
Let us write the Dyson equation at $T=0$ in the form
\begin{align}
\G & = \G^0 + \G^0 \Sigma \G \,  = \sum_{n=0}^{\infty} \G^0 \left\{\Sigma \G^0\right\}^n ,
\label{eq:DYSO}
\end{align}
where $\G^0$ is the \emph{time-ordered} noninteracting sp propagator.
The momentum distribution can then be written as
\begin{align}
n(k) & = \int_{C} \frac{\textrm{d} \w}{2\pi i} \G(k,\w) 
\label{eq:momself} \\ 
     &=\int_{C} \frac{\textrm{d} \w}{2\pi i} \sum_{n=0}^{\infty} \G^0 \left\{\Sigma \G^0\right\}^n .
\nonumber
\end{align}
with $C$ representing a contour composed of the real axis and closed by a semi-circle in the
upper half-plane.
In order to proceed it is necessary to decompose the self-energy into its different time-ordered
contributions
\begin{align}
\Sigma(k,\w ) & = \Sigma_{HF}(k) + \Sigma_\downarrow(k,\w) + \Sigma_\uparrow(k,\w) = 
\label{eq:selfud} \\
\Sigma_{HF}(k) - & \int_{\e_F}^\infty \!\! \frac{\textrm{d} \w'}{\pi} \frac{\textrm{Im} 
\Sigma_\downarrow(k,\w')}{\w_+ - \w'} +
 \int_{-\infty}^{\e_F} \!\! \frac{\textrm{d} \w'}{\pi} \frac{\textrm{Im} 
\Sigma_\uparrow(k,\w')}{\w_- - \w'} . \nonumber
\end{align}
The first term in the expansion of Eq.~(\ref{eq:momself}),
corresponding to $\G^0$, yields the step function momentum distribution,
$n(k)^0=\Theta (k_F- k)$, of the FFG.
Using Eq.~(\ref{eq:selfud}), the second term ($n=1$) in Eq.~(\ref{eq:momself}) yields three 
contributions, each of them with two noninteracting propagators, $\G^0$.
For $k < k_F$, the contribution with the Hartree-Fock term vanishes on 
account of the double pole inside the contour, which leads to the
derivative of a constant when Cauchy's integral formula is applied.
The contribution with $\Sigma_\uparrow$ can be shown to be zero as
well by using the corresponding dispersion relation for
$\Sigma_\uparrow$ and applying the residue theorem (the residues cancel).
Only the term with $\Sigma_\downarrow$ finally contributes and yields:
\begin{align}
n_\uparrow^1(k) = \partial_\w \textrm{Re} \Sigma_\downarrow(k,\w) |_{\w=\e^0(k)} \, , \,\,\, 
k < k_F \, , \label{eq:n1b}
\end{align}
by applying Cauchy's integral formula. The derivative is applied at the pole of the 
noninteracting propagator.
Since the peak of the dressed propagator will normally occur at the quasi-particle energy
\begin{align}
\e(k) = \frac{k^2}{2m} + \textrm{Re} \Sigma(k,\e(k)) , \label{eq:eqp}
\end{align}
it is convenient to include an auxiliary potential
\begin{align}
U(k) = \textrm{Re} \Sigma(k,\e(k)) , \label{eq:uk}
\end{align}
anticipating that the derivative in Eq.~(\ref{eq:n1b}) must be taken
at the location where most of the sp strength is concentrated.
This leads to an additional constant term in Eq.~(\ref{eq:selfud}), where $U$ must be 
subtracted.
An approximate result for $n(k)$ for $k<k_F$ can now be obtained by expanding 
$\Sigma_\downarrow$ only up to first order at $\e(k)$ in all contributions to 
higher-order terms ($n>1$) in Eq.~(\ref{eq:momself}).
This approximation will therefore be best for momenta as far as possible from the Fermi energy,
since from the corresponding dispersion relation it is clear that a linear approximation to the
energy dependence will be most appropriate for $k=0$ (for which $\e(0)$ is farthest from $\e_F$).
The series with only $\Sigma_\downarrow$ terms yields
\begin{align}
n_\uparrow(k) = \frac{1}{1 - \partial_\w \textrm{Re} \Sigma_\downarrow(k,\w) |_{\w=\e(k)} } \, , \,\,\, k < k_F \, , 
\label{eq:momhole}
\end{align}
since a geometric series is produced with the energy derivative appearing in each higher order.
Other terms including the constant $\Sigma_{HF}$ and $\Sigma_\uparrow$ will not contribute in a 
similar fashion as in the term with $n=1$, as long as the validity of this expansion for 
$\Sigma_\downarrow$ is assumed.

For momenta above $k_F$, the noninteracting propagators in
Eq.~(\ref{eq:momself}) have poles outside the contour $C$ and cannot contribute.
Only terms with the derivative of $\Sigma_\uparrow$ will occur when a similar
expansion as for $k < k_F$ is employed.
The $n=1$ term accordingly generates the first contribution to $n(k)$ for $k> k_F$, given by
\begin{align}
n^1_\downarrow(k) = -\partial_\w \textrm{Re} \Sigma_\uparrow(k,\w) |_{\w=\e(k)} \, , \,\,\, k > k_F \, .
\label{eq:mompart}
\end{align}
For high momenta this will be the dominant contribution to $n(k)$, since the location of $\e(k)$
can be quite far from $\e_F$.
Summing the geometric series from higher-order contributions with $\Sigma_\uparrow$ (with the
approximation of keeping only linear terms in the energy) then yields
\begin{align}
n_\downarrow(k) =1- \frac{1}{1 - \partial_\w \textrm{Re} \Sigma_\uparrow(k,\w) |_{\w=\e(k)} } \, , \,\,\, k > k_F \, . 
\label{eq:momparta}
\end{align}
While Eqs.~(\ref{eq:momhole}) and (\ref{eq:momparta}) have not been rigorously derived here, 
we note that these results have been employed before in the literature~\cite{baldo92}.
Indeed, it turns out that their numerical implementation generates an accurate approximation to 
the full result for $n(k)$, except in the immediate vicinity of the Fermi surface, as will 
be illustrated in the following.
Moreover, these equations allow for a deeper understanding of the observed trends in the depletion
of the Fermi sea.
Since the decomposition of the self-energy in terms of ``up'' and
``down'' components is essential for this derivation,
it clarifies the need for introducing these terms also at finite temperature.

\begin{figure}[t]
  \begin{center}
    \includegraphics*[width=\linewidth]{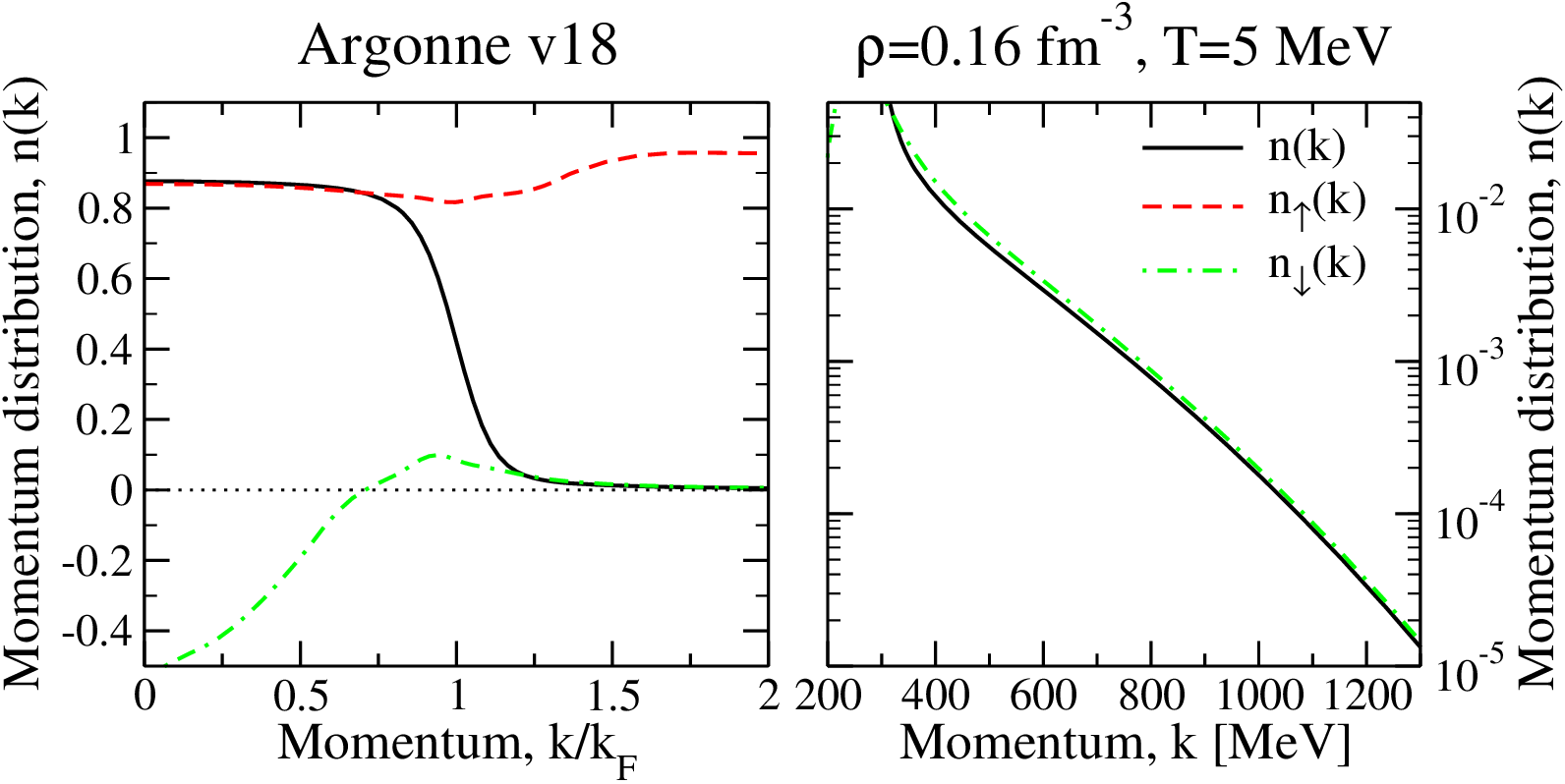}  
    \caption{(Color online) Correlated momentum distribution obtained with the Av18 interaction at $\rho=0.16$ fm$^{-3}$ and $T=5$ MeV (solid line). The $n_\uparrow(k)$ and $n_\downarrow(k)$ contributions of Eqs.(\ref{eq:momhole}) and (\ref{eq:momparta}) are shown in dashed and dot-dashed lines, respectively.}
    \label{fig:momupdw}
  \end{center}
\end{figure}

For a self-consistent calculation at finite temperature, expressions~(\ref{eq:momhole}) and 
(\ref{eq:momparta}) reproduce the momentum distribution obtained from the complete spectral 
function extremely well in their respective momentum range. 
An example is shown in Fig.~\ref{fig:momupdw} for Av18 at $\rho=0.16$ fm$^{-3}$ and $T=5$ MeV. 
The hole contribution coming from Eq.~(\ref{eq:momhole}) (dashed line,
left panel) is a very good approximation to the 
momentum distribution for momenta below $0.75 k_F$. 
Similarly, for momenta above $1.25 k_F$, $n(k)$ is very well approximated by 
Eq.~(\ref{eq:momparta}) (dash-dotted line, right and left panels). 
A similar agreement is found in a wide range of densities and
temperatures, and also for other NN forces. 
We note that, for these results to be valid, it is essential to calculate the derivatives at the corresponding quasi-particle
energies and therefore self-consistency is a necessary ingredient in obtaining this good
agreement.

\begin{figure}[t!]
  \begin{center}
    \includegraphics*[width=0.75\linewidth]{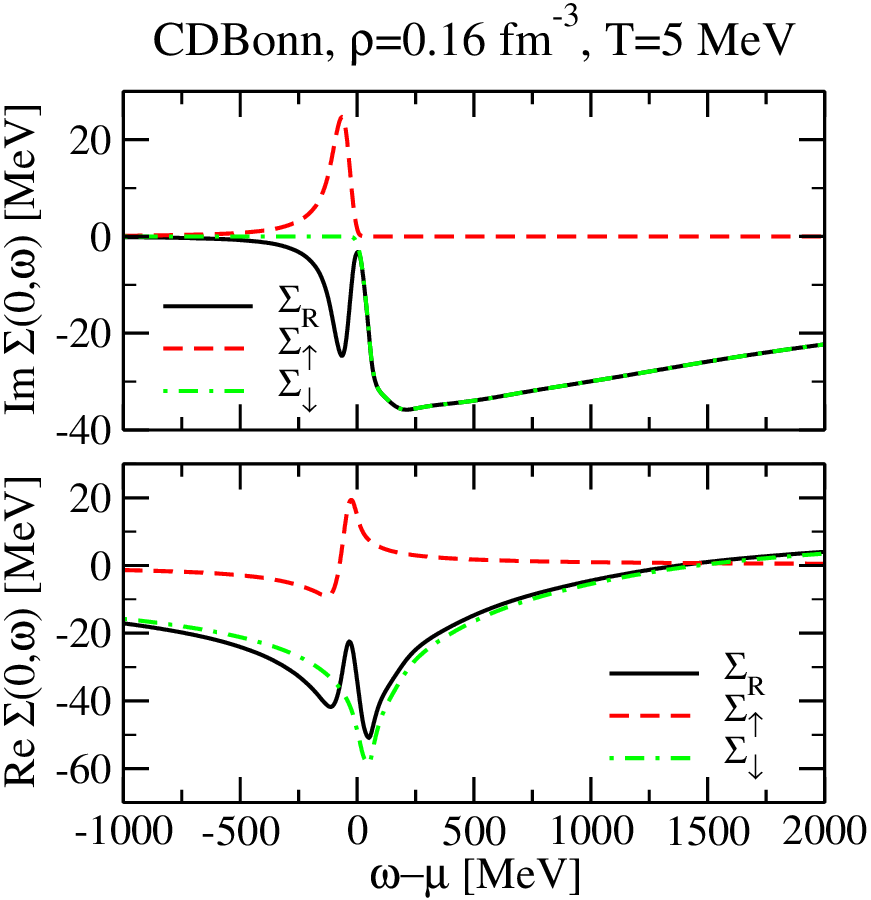}  
    \caption{(Color online) Different components to the imaginary (upper panel) and real (lower 
panel) parts of the self-energy for CDBonn at $\rho=0.16$ fm$^{-3}$ and $T=5$ MeV. }
    \label{fig:selfdec}
  \end{center}
\end{figure}
Having found a direct connection between $n(k)$ and the self-energy, one can try to establish 
a relation between the density dependences of $n(k)$ and the density and energy dependence of the 
self-energy. 
To this end, let us discuss the structure of $\Sigma_\uparrow$ and $\Sigma_\downarrow$. 
Figure~\ref{fig:selfdec} shows the self-energy of the $k=0$ state as a function of energy for the 
CDBonn potential at $\rho=0.16$ fm$^{-3}$ and $T=5$ MeV. 
The upper panel displays three different components of $\Sigma$. 
The solid line corresponds to the retarded self-energy, which has traditionally been used and 
discussed in finite-temperature SCGF calculations~\cite{schnell96,frickphd}. 
Note that $\textrm{Im} \Sigma_\R$ is negative at all energies and displays two distinctive peaks: 
a relatively small peak in the ``hole'' region, $\w < \mu$, and a much larger structure in 
the ``particle'' region, $\w > \mu$. 
While the hole components decay rapidly for negative energies, the particle side has a much 
slower decay in energy due to the combination of the large particle-particle phase-space and 
short-range correlation effects, in complete analogy with earlier results obtained for 
$T=0$~\cite{ramos89}.

Since the self-energy has the same analytical structure as the sp propagator, the time-ordered 
contributions of $\Sigma$ can be obtained from the retarded component by using equivalent 
expressions to Eqs.~(\ref{eq:up_r}) and (\ref{eq:down_r}). 
The phase space factors in the imaginary parts isolate the respective particle and hole regions. 
Consequently, $\textrm{Im} \Sigma_\uparrow$ and $\textrm{Im} \Sigma_\downarrow$ have a single 
peak below and above the Fermi surface, respectively. 
In the case of the hole contribution, the peak has the opposite sign compared to the retarded 
component, but for the particle one it coincides with the 
$\textrm{Im} \Sigma_\R$ peak. 
The differences between the retarded and the corresponding time-ordered imaginary components 
will be larger in the $\w \sim \mu$ region, due to the variation of the thermal distribution in 
this region. 
In particular, at higher temperatures, the opening of phase-space can eventually lead to the appearance of additional peaks in the self-energy of each component, due to the overlap between the ``opposite'' peak and the non-zero Fermi-Dirac distribution. 

The different imaginary components of $\Sigma$ are related to their respective real parts by the 
dispersion relations of Eqs.~(\ref{eq:dispersion_R}), (\ref{eq:dispersion_up}) 
and (\ref{eq:dispersion_down}).  
These real parts are illustrated in the lower panel of Fig.~\ref{fig:selfdec}. 
The important consequences of the isolation of the hole and particle components in the imaginary 
part can be clearly observed now.
$\textrm{Re} \Sigma_\uparrow$ decays rapidly at low and high 
energies, because the hole strength in the imaginary part is limited to a relatively narrow 
energy domain. 
This component exhibits a wiggle at small negative energies, associated with the 
hole peak. 
The down component of $\Sigma$ has a substantial contribution at all energies, as a consequence 
of the extremely long energy tail of the imaginary part~\cite{frickphd,riosphd,dickhoff05}. 
It also has a peak structure, related to the onset of a very fast decrease of 
$\textrm{Im} \Sigma_\downarrow$ near $\w \sim \mu$. 
This threshold, together with the relatively large and constant energy tail, are at the origin of 
a bell-shaped minimum. 
As we shall see in the following, it is this minimum that dominates the low-energy components of 
the momentum distribution. 
The real part of the down component has a zero at high energies, $\sim 1.5$ GeV, and it is still 
substantial at very high energies.
These results are in good qualitative agreement with the zero-temperature calculations of 
Ref.~\cite{ramos89} for a much harder interaction. 

The decomposition of $\Sigma$ in its time-ordered components can also be used to better understand the 
structure of the retarded self-energy. 
The sum of $\textrm{Re} \Sigma_\uparrow$ and $\textrm{Re} \Sigma_\downarrow$ must coincide with 
the real part of the retarded self-energy \cite{mahan}. 
The high positive and negative energy tails of the retarded
self-energy have to be related to the down component of $\Sigma$,
which is the only active contribution in these regions. 
Moreover, the double-peak structure of $\textrm{Re} \Sigma_\R$ at $\w \sim \mu$ is 
caused by the superposition of the up and down peaks and wiggles in this region. 
In particular, the $\w < \mu$ peak in $\textrm{Re} \Sigma_\R$ is a reflection of the negative side of the wiggle in 
$\textrm{Re} \Sigma_\uparrow$, while the $\w > \mu$ peak is mostly due to the down component. 
The relative importance of the hole and particle peaks changes with momentum, with 
the ``particle'' side becoming more prominent as $k$ increases. 
In spite of this important shift, the connection between the up and down components and the high 
energy and peak structures of the retarded self-energy remains valid. 

\begin{figure}[t!]
  \begin{center}
    \includegraphics*[width=\linewidth]{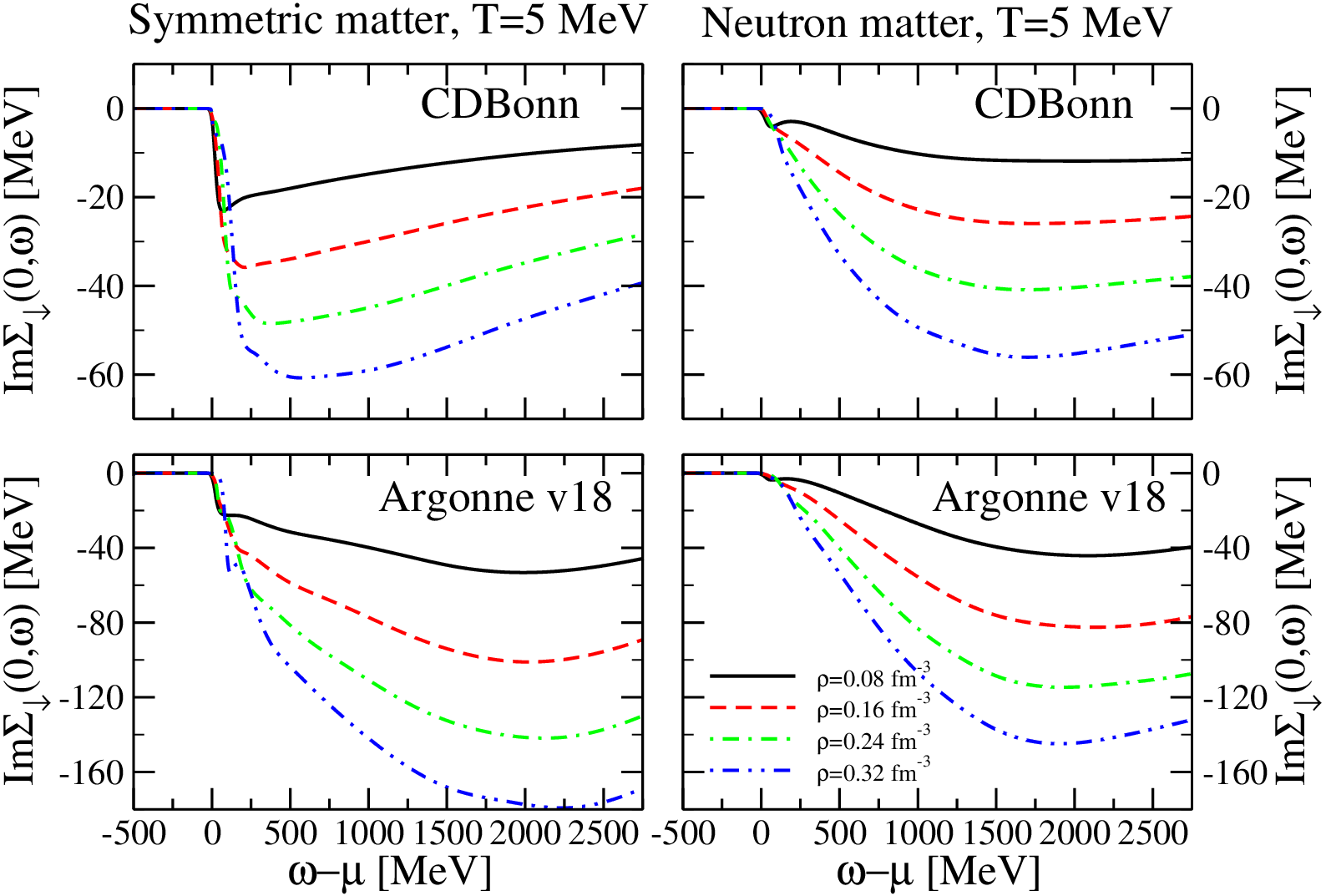}  
    \caption{(Color online) Imaginary part of $\Sigma_\downarrow$ as a function of energy for the $k=0$ state. Different densities are displayed with different line styles. The upper panels have been obtained with CDBonn, while the lower panels correspond to Av18. Symmetric (neutron) matter results are shown in the left (right) panels. Note the difference in scales between the upper and lower panels.}
    \label{fig:imselfk0}
  \end{center}
\end{figure}

The $\Sigma_\downarrow$ self-energy is related, via Eq.~(\ref{eq:momhole}), to the momentum 
distribution below $k_F$. 
This suggests that the density dependence of the low-momentum components of $n(k)$ can be 
understood in terms of the self-energy. 
Let us focus for simplicity on the $k=0$ state. 
Figure~\ref{fig:imselfk0} displays the energy dependence of $\textrm{Im} \Sigma_\downarrow$ for 
different densities in the case of symmetric (left panels) and neutron matter (right panels). 
The upper (lower) panels correspond to the CDBonn (Av18) interactions. 
The imaginary part of $\Sigma_\downarrow$ has a very large structure in the $\w > \mu$ region 
that only decays at extremely high energies. 
The area covered by this function depends substantially on the density, the NN potential and the 
isospin content of the system. 
In the high energy regime, $\textrm{Im} \Sigma_\downarrow$ displays a linear dependence in density 
for both nuclear and neutron matter. 

A significant difference is found by comparing upper and lower panels in Fig.~\ref{fig:imselfk0}: 
the amount of imaginary part shifted to high energies is substantially larger for Av18 than for 
CDBonn (note the very different vertical scales). 
The down component of the self-energy is therefore very sensitive to the structure of the 
underlying NN force. 
This result suggests that hard interactions produce a much larger imaginary part than soft ones 
do~\cite{dickhoff05}, for both SNM and PNM. For a given NN
interaction, however, the peak of $\textrm{Im} \Sigma_\downarrow$ at high energies is 
relatively similar for the two systems, as seen by comparing left and right 
panels. 
While in PNM the peak can be attributed basically to short-range
correlations, SNM 
has an additional contribution coming from the tensor components that generates an important
imaginary part below 1000 MeV, as discussed in Refs.~\cite{Vond91,Vond93}. 
It is good to remember that the presence of an imaginary part in the self-energy above the 
chemical potential is directly responsible for the appearance of sp strength at those energies
for momenta below $k_F$. 
In nuclear matter, the additional ${}^3S_1$-${}^3D_1$ coupled channel with both its short-range 
and tensor contribution is thus responsible for the stronger and more pronounced imaginary part 
starting above the chemical potential as compared to neutron matter, where it is absent. 
The position of the peak in $\textrm{Im} \Sigma_\downarrow$ also changes when going from 
symmetric to neutron matter, but in an interaction-dependent way. 
While for CDBonn the peak is shifted to high energies for neutron matter, the contrary happens 
for Av18~\cite{dickhoff05}. 

\begin{figure}[t!]
  \begin{center}
    \includegraphics*[width=\linewidth]{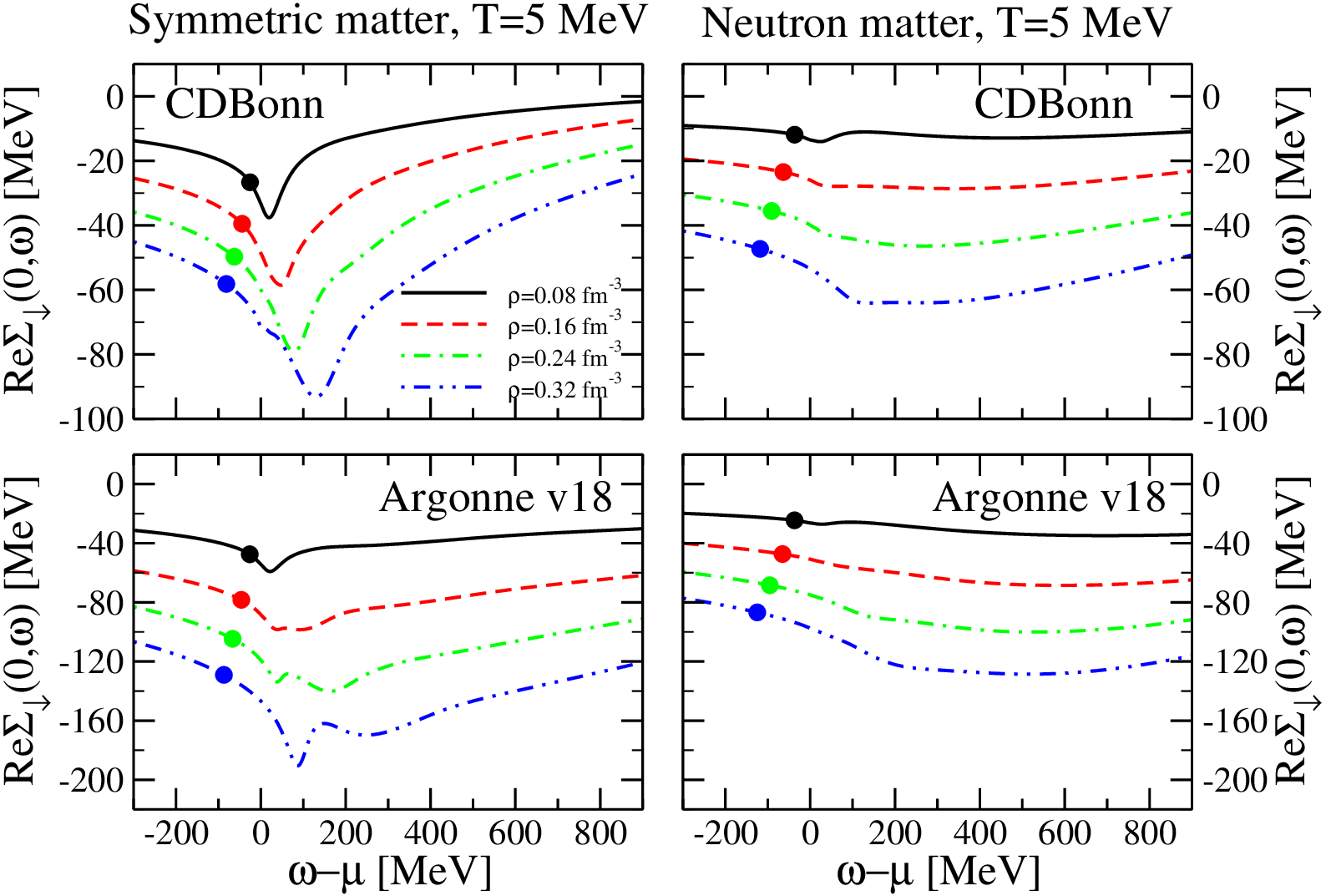}  
    \caption{(Color online) The same as Fig.~\ref{fig:imselfk0} for the real part of $\Sigma_\downarrow$. The dots represent the position of the quasi-particle peak. }
    \label{fig:reselfk0}
  \end{center}
\end{figure}

The energy dependence of the imaginary part has an immediate impact on the real part of 
$\Sigma_\downarrow$ by means of the dispersion relation. 
Figure~\ref{fig:reselfk0} shows this component in a narrower energy range for various densities. 
In general, the real part is proportional to the density, with higher densities leading to 
more attractive real self-energies (in the range displayed in the figure). 
Comparing left and right panels, one observes an important difference between symmetric and 
neutron matter. 
While a distinctive inverted peak develops in nuclear matter for the two NN interactions, in 
neutron matter only a very broad minimum is observed. 
The origin of this peak is associated with the rapid drop, from zero to negative values, of 
$\textrm{Im} \Sigma_\downarrow$ in a narrow energy band around $\w \sim \mu$. 
This sharp threshold occurs for symmetric matter (see the upper left panel of 
Fig.~\ref{fig:imselfk0}, for instance), but it is not as sharp in neutron matter, reflecting
the important on/off role of the tensor force in these systems.
As a consequence of these large differences, the dispersive counterparts of the threshold of 
the imaginary part are very different for both systems.

The presence of the inverted peak in symmetric matter has important consequences for the momentum 
distribution at low momenta. 
To compute $n(k)$, the partial derivative with respect to the energy of 
$\textrm{Re} \Sigma_\downarrow$ is computed at the quasi-particle peak. 
The quasi-particle energy for $k=0$ at different densities is identified by dots in 
Fig.\ref{fig:reselfk0}. 
As expected, at zero momentum the quasi-particles are more bound for higher densities, so their 
energies shift to more negative values with respect to the chemical potential. 
In SNM, this shift moves the quasi-particle contribution further away
from the inverted peak, which is displaced to higher energies as density increases. 
The combination of the two effects (attractive quasi-particle shift, repulsive peak shift) is 
such that the slope at the quasi-particle energy is reduced with increasing density. 
For PNM, the quasi-particle shift is larger, but the inverted peak is absent
and the broad structure is such that the local slope around the quasi-particle contribution 
decreases with density. 

\begin{figure}[t!]
  \begin{center}
    \includegraphics*[width=\linewidth]{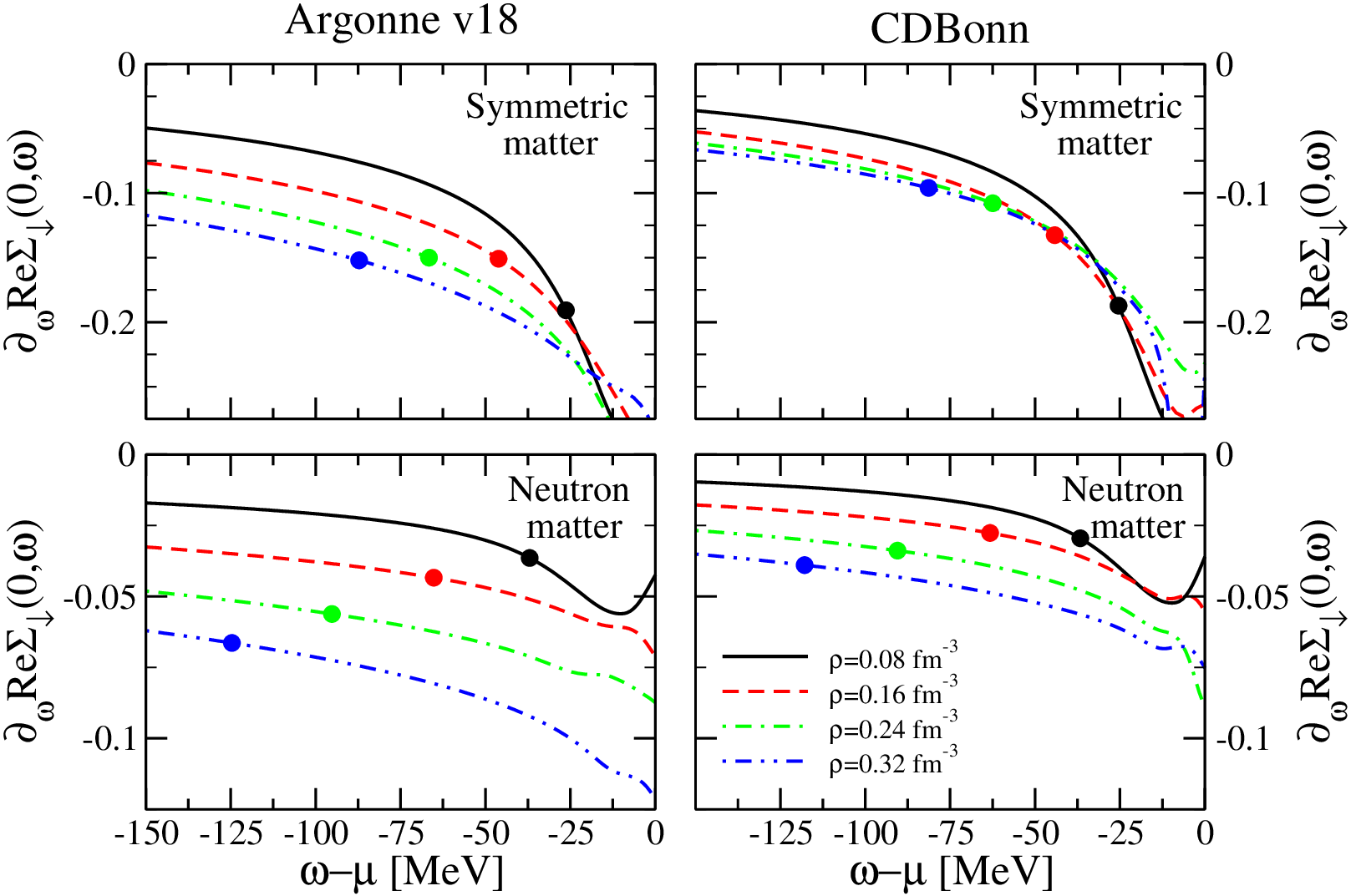}  
    \caption{(Color online) Partial derivative of $\Sigma_\downarrow$ as a function of energy for the $k=0$ state at $T=5$ MeV. Different densities are displayed with different line styles. The upper panels show symmetric matter results, while the lower panels correspond to neutron matter. Left (right) panels have been obtained with the Av18 (CDBonn) interaction. }
    \label{fig:derselfk0}
  \end{center}
\end{figure}

To clarify these issues, we show in Fig.~\ref{fig:derselfk0} the partial derivative with respect 
to the energy of the self-energy at $T=5$ MeV for different densities. 
Upper (lower) panels correspond to symmetric (neutron) matter, while left (right) panels show 
results for Av18 (CDBonn). 
Let us first focus on the SNM case (upper panels). 
At all densities, the inverted peak in the real part (a zero in the derivative) lies inside the $\w > \mu$ 
region, and it is therefore not visible in the figure. Instead, we
have chosen to display the $\w < \mu$ regime, 
where the derivative should be taken according to Eq.~(\ref{eq:momhole}). 
This corresponds to the wings of the peak on the left side. 
At these energies, the slope of $\textrm{Re} \Sigma_\downarrow$
decreases with energy and, in general, has more negative values at higher densities.  
The increase of the slope with density also appears to be related to the hardness of the interaction. 
While for CDBonn the slope changes very little as density increases, Av18 leads to much larger 
variations in SNM. 
The soft interaction, therefore, generates a small variation in density of 
$\partial_\w \textrm{Re} \Sigma_\downarrow$ which, together with the quasi-particle shift, 
produces a \emph{decreasing} depletion as the density increases. 
This observation therefore explains the counter-intuitive behavior observed in the left panel of 
Fig.~\ref{fig:depsymneu}. 
For a harder interaction, such as Av18, the effect of density in the slope is much larger, 
while the quasi-particle shift is similar to CDBonn. 
Consequently, there is the possibility for the depletion to saturate (or even decrease) at high 
densities. 
Earlier results for a separable version of the Paris potential exhibit little density 
dependence for the depletion~\cite{baldo90} not unlike our results for Av18 in the range
from once to twice normal density.
Note that the density dependence of the self-energy is closely related to the strong correlation 
effects induced by the short-range components. 

The density dependence of the  partial derivative is also important in neutron matter (lower 
panels). 
Again, the slope changes much more as a function of density for Av18 than for CDBonn. 
Due to the quasi-particle shift, the derivative at zero momentum decreases. 
Thus, the depletion increases with density, in agreement with the right panel of 
Fig.~\ref{fig:depsymneu}. 
For a given NN interaction, the overall decrease of $\partial_\w \textrm{Re} \Sigma_\downarrow$ 
with density and the quasi-particle shift are qualitatively similar for both neutron and nuclear 
matter. 
The origin of the differences between SNM and PNM should thus be attributed to the energy dependence of the slope, 
which is much more pronounced for symmetric matter than for neutron matter. 
This is particularly true for energies close to $\w \sim \mu$, where a sharp decrease is observed 
associated with the inverted peak in the real self-energy. 
As we have already discussed, this is largely related to the threshold behavior of the imaginary 
part of $\Sigma_\downarrow$. 
In turn, this behavior must be partly ascribed to the tensor component of the 
one-pion-exchange interaction.


\section{Momentum distribution of two-component nuclear systems\label{sec:mom2}}

Up to now, we have considered the momentum distributions for two
extreme cases, namely SNM and PNM.
It is also illustrative  to study the dependence 
of the proton and neutron momentum distributions, $n_{\tau}(k)$, 
as a function of the asymmetry parameter, $\alpha$. 
This analysis should shed some light on the the isospin dependence of single-particle nuclear
properties coming from volume effects. 
In particular, tensor effects can be highlighted in asymmetric systems
and can be pinpointed by a comparison of different NN potentials.

\begin{figure}[t]
  \begin{center}
    \includegraphics*[width=\linewidth]{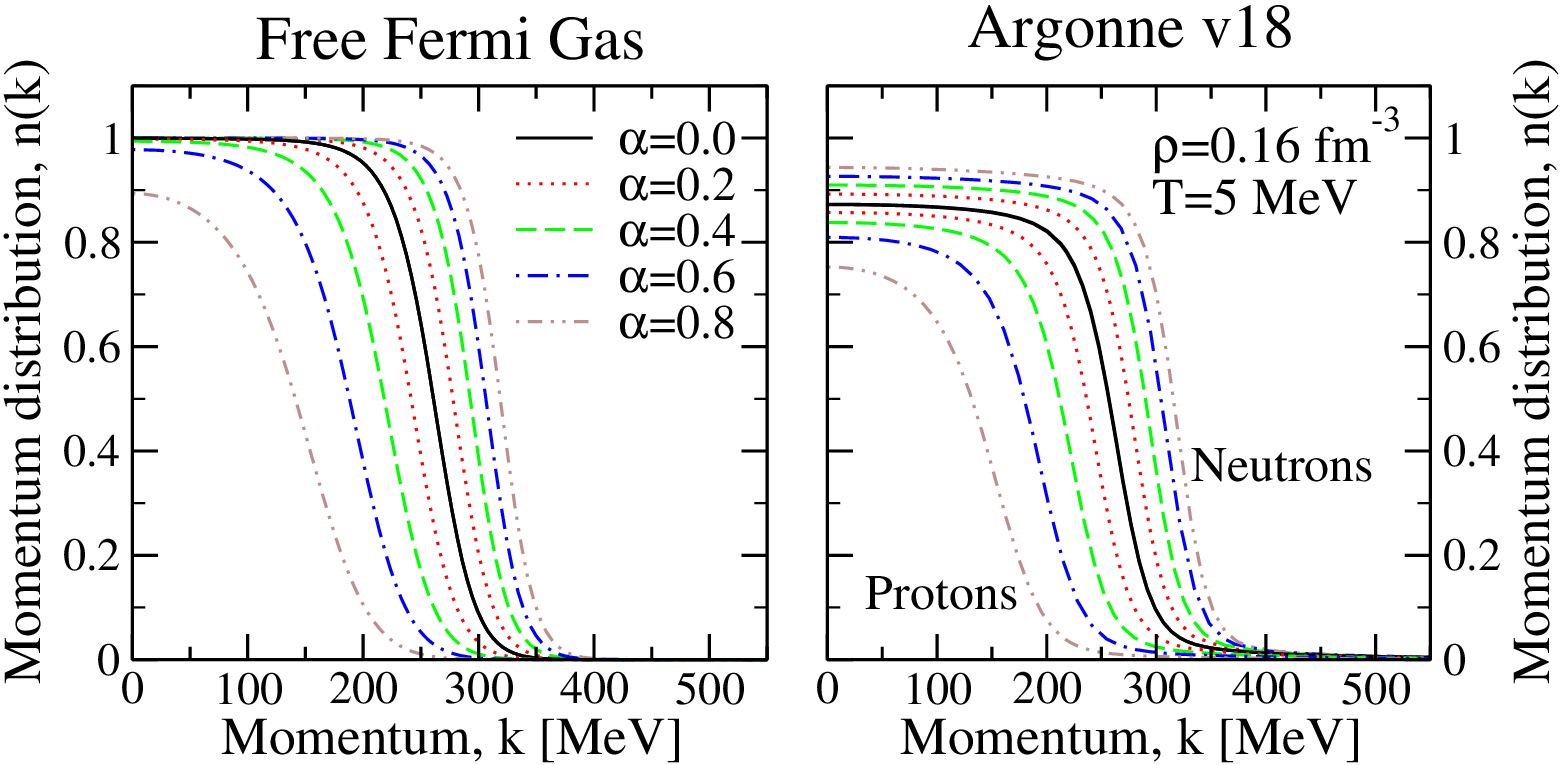}  
    \caption{(Color online) Isospin asymmetry dependence of the Free Fermi gas (left panel) and the correlated momentum distribution from Av18 (right panel). The density ($\rho=0.16$ fm$^{-3}$) and temperature ($T=5$ MeV) are fixed and the results for different asymmetries are shown in different line-styles. Symmetric matter results correspond to the central solid curve, while curves to its right (left) correspond to neutron (proton) distributions for the corresponding isospin asymmetric cases.}
    \label{fig:qp}
  \end{center}
\end{figure}

The asymmetry dependence of the neutron and proton momentum
distribution is presented in Fig.~\ref{fig:qp} 
at $\rho=0.16$ fm$^{-3}$ and $T=5$ MeV. 
The left (right) panel contains the results for FFG (Av18). 
For $\alpha=0$ (symmetric case), the momentum distributions of
neutrons and protons coincide in both cases.
Let us discuss first the left panel,
where we show the momentum distribution of 
the isospin asymmetric FFG.
For the symmetric case ($\alpha=0$) and as discussed previously in
the context of Fig.~\ref{fig:momsym}, the thermal effects on $n(k)$ 
are only visible at the Fermi surface and almost no depletion occurs 
inside the Fermi sea. 
The same is true for the most abundant component as asymmetry
increases: the depletion is negligible and the Fermi momentum is
displaced to larger values. For protons, however,
$n(k)$ is clearly affected by temperature, exhibiting a large 
change in shape in a wide region of momenta. At large asymmetries, in
particular, the momentum
distribution of the less abundant components differs substantially from the step 
function, loosing its degeneracy character and decreasing the occupation at the origin.  
In some sense, when increasing the asymmetry, the less abundant component 
moves closer to a classical momentum distribution, while the most
abundant component becomes more degenerate. 
Note however that, at $\alpha= 0.8$, one is still far away from the
classical regime even though thermal effects are important.
This analysis suggests that one should take into account thermal
corrections when analyzing the change in asymmetry of $n_\tau(k)$ also
for the correlated case.

Focusing on the right panel, one observes that the most abundant component (neutrons for 
positive $\alpha$) gets less depleted when the asymmetry increases,
\emph{i.e.} neutrons become ``less correlated''. This is in contrast
to the FFG results, which, for neutrons, exhibit no change inside the
Fermi sea. The decrease of depletion for the most abundant component
in this density and temperature should be taken as a pure NN
correlation effect. This behavior can be explained in simple terms as follows.
Although the total number of pairs is the same when increasing the asymmetry at constant 
density, some of the pn pairs are replaced by neutron-neutron pairs. The latter
correlations are weaker than the pn ones, due to the absence of tensor
effects, and therefore neutrons become less correlated at large $\alpha$'s.
Conversely, the momentum distributions of the less abundant species
(protons) become more depleted with asymmetry. 
A single proton sees an increasing number of neutrons when the asymmetry increases, \emph{i.e.} 
pp pairs are replaced by the more correlated pn pairs, which results
into a more depleted proton momentum distributions. Together with
this, protons ``feel'' more the effect of temperature. 
The proton density in the asymmetric system decreases as $\alpha$
increases. At finite temperature, this is translated into the fact
that $n_p(k)$ becomes dominated by thermal effects at large
asymmetries and looks less degenerate. The plateau associated with particle
states, for instance, which is clearly observed for neutrons at all asymmetries, gets
washed out for protons at large $\alpha$. 
Consequently, one should be careful in associating the increase of the 
depletion of protons only to 
dynamical NN correlation, because the momentum distribution of the
less abundant component might be strongly influenced by thermal effects. 

\begin{figure}[t]
  \begin{center}
    \includegraphics*[width=\linewidth]{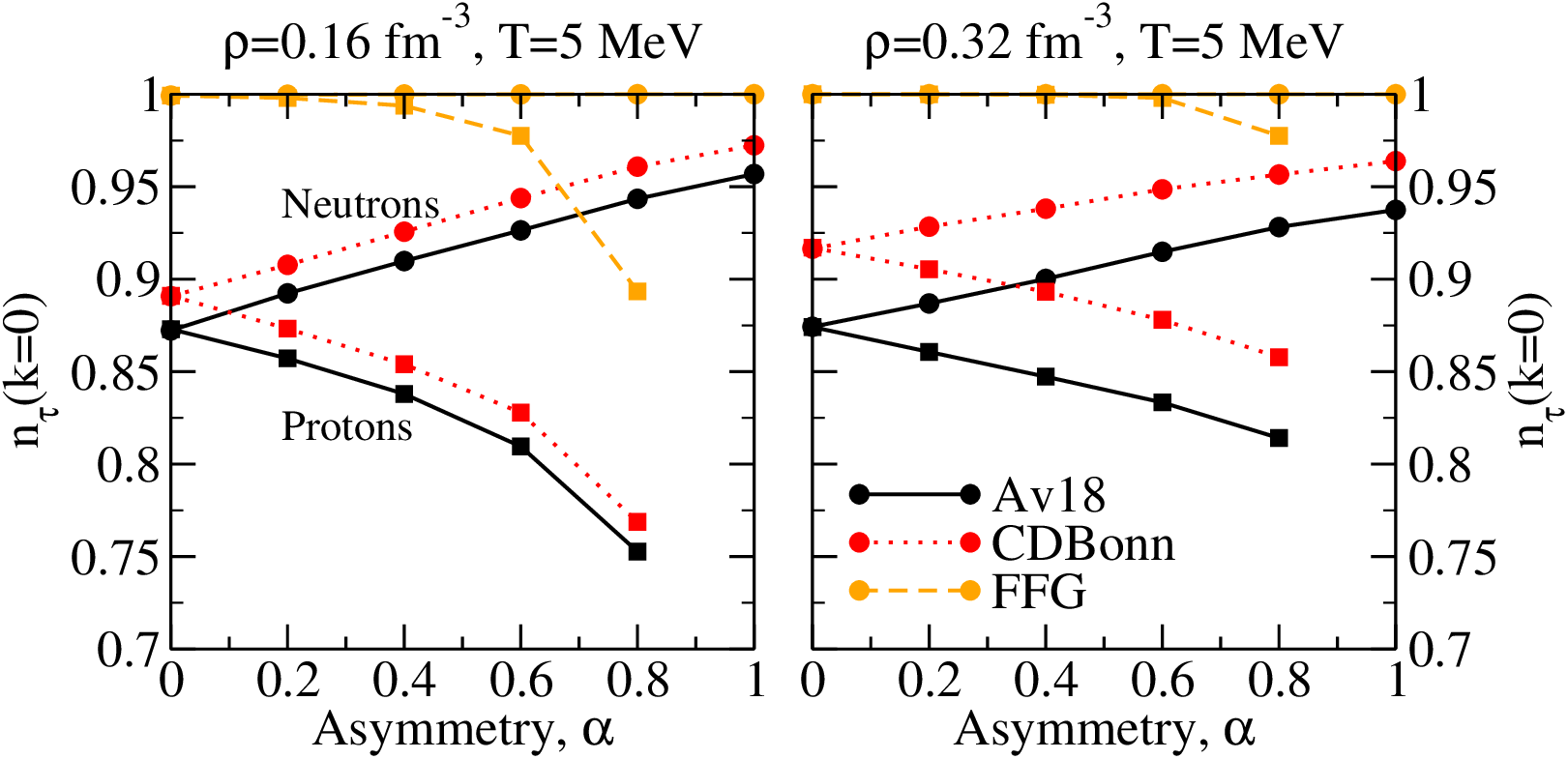}  
    \caption{(Color online) Isospin asymmetry dependence of the neutron (filled circles) and proton (squares) lowest momentum occupation. Correlated results for the Av18 (solid lines) and CDBonn (dotted lines) interactions are compared to the Free Fermi Gas (dashed lines) predictions.}
    \label{fig:depasy}
  \end{center}
\end{figure}

The information about the isospin dependence of the 
depletion is summarized in Fig.~\ref{fig:depasy}, where  
$n_{\tau}(0)$ is plotted as a function of the asymmetry at $T=5$ MeV for $\rho=0.16$ fm$^{-3}$ 
(left panel) and $\rho=0.32$ fm$^{-3}$ (right panel). 
The FFG (dashed lines) results are compared to the Av18 (solid lines)
and CDBonn (dotted lines) SCGF predictions. 
Again, the FFG results provide a measure of thermal effects. 
On the one hand, the occupation of the most abundant component at zero momentum 
does not change when the asymmetry 
increases, indicating that this species is totally degenerate in the
whole range of asymmetries. On the other hand, the corresponding
occupation of protons (the less abundant component) is close to $1$ at
small asymmetries and decreases as the asymmetry increases. At
$\rho=0.16$ fm$^{-3}$, this accounts for a $\sim 2 \%$ effect for
$\alpha=0.6$ and a $\sim 11 \%$ for $\alpha=0.8$. As a matter of fact,
in the limit $\alpha \to 1$, the protons become an impurity gas in a
Fermi sea of neutrons, thus behaving as a classical gas with
$n_p(0) \to 0$. The steepness of the change in $n_p(0)$ depends 
on the total density and the temperature of the system. At a higher
density (right panel), the thermal effects affecting protons 
do not set in unless the asymmetry is very large 
and, even at $\alpha=0.8$, the effect is only of $\sim 2 \%$. 
Let us stress the fact that this is a pure thermal effect: for the FFG at zero temperature, 
$n_{\tau}(0)=1$ for both protons and neutrons at all asymmetries.

For the correlated depletion, the occupation of the zero momentum state is an 
increasing (decreasing) function of the asymmetry for the most (less) abundant component. 
The behavior is very similar for both NN interactions, although the occupations for Av18 are 
systematically smaller than those for CDBonn at both densities. 
The differences between the two potentials increase with density (see
right panel). 
The comparison with the FFG allows us to identify the regime in which NN
correlations dominate over thermal effects. At $T=5$ MeV and
$\rho=0.16$ fm$^{-3}$, for instance, and up to asymmetries of about 
$\alpha \sim 0.4-0.6$, the occupations 
of neutrons and protons can be attributed to dynamical correlations and should provide a 
good estimate for the case of zero temperature. Up to these
asymmetries, both the neutron and the proton depletion change almost
linearly with asymmetry. In the right panel,  the depletion of
protons starts to bend down for larger asymmetries, attributable to the onset of thermal
effects. In a more degenerate case (right panel), where thermal effects
are almost negligible, the asymmetry dependence of
both $n_\tau(0)$'s is again found to be linear. 

Asymmetries of stable nuclei belong to the lower range of asymmetries, 
\emph{i.e.} $\alpha \sim 0.2$ for $^{208}$Pb. The SCGF predictions
should be valid in this range, where thermal effects are unimportant. 
Unfortunately, the difference between the occupation of protons at 
this asymmetry and the symmetric 
case is only $\sim 2 \%$, too small to allow experimental verification.
Nuclei at larger asymmetry produced at future rare isotope facilities may provide a better 
testing ground. Moreover, integrated effects over the whole neutron
and proton density profiles might also enhance the effect of asymmetry. 
Let us note that, for finite nuclei, one should also take into account
surface properties and their isospin dependence. In particular, an
asymmetry dependent effect associated with surface properties has been 
identified for protons in Ca isotopes~\cite{charity06,charity07}. 

\begin{figure}[t]
  \begin{center}
    \includegraphics*[width=\linewidth]{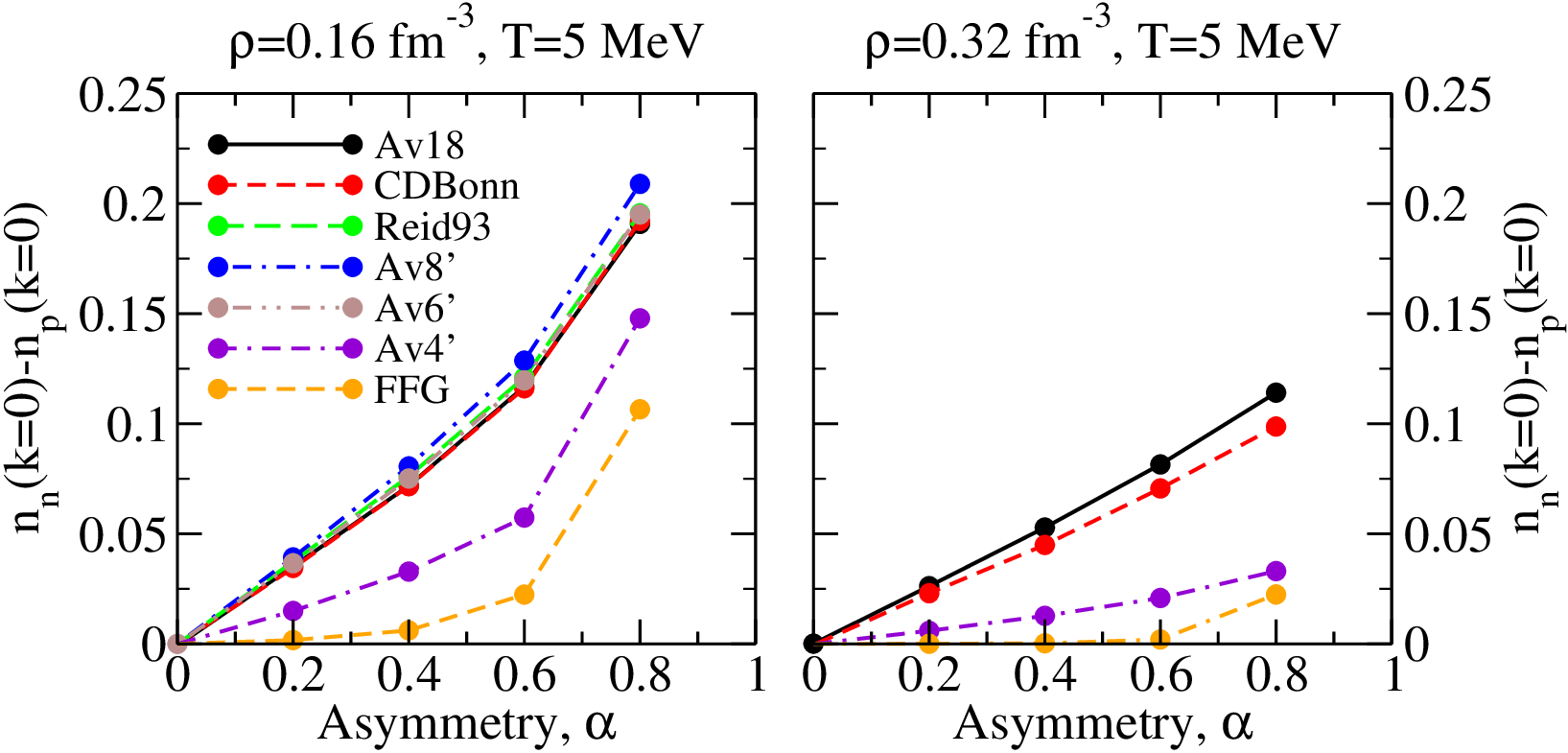}  
    \caption{(Color online) Difference of neutron and proton
      occupation of the lowest momentum state as a function of isospin
      asymmetry for different interactions. The two panels show
      results at $T=5$ MeV for $\rho=0.16$ fm$^{-3}$ (left) and
      $\rho=0.32$ fm$^{-3}$ (right).}
    \label{fig:depl}
  \end{center}
\end{figure}

A surprising feature arises when comparing the two panels of
Fig.~\ref{fig:depasy}. In spite of the differences observed in both
SNM and PNM for the Av18 and CDBonn results, the asymmetry dependence
of $n_\tau(0)$ is very similar at all densities. This is
shocking because, as we have stressed so far, both forces have a
rather different short-range behavior and tensor structure. Given the
almost linear dependence of $n_\tau(0)$ with asymmetry, a better
insight into these differences can be gained by plotting the difference 
$n_n(0)-n_p(0)$. In Fig.~\ref{fig:depl}, we present this
``iso-depletion'' for different interactions as a function of the 
asymmetry at $T=5$ MeV for the same densities as before. 
The left panel shows that this difference is 
the same for a wide variety of modern NN potentials, independently of their
short-range or operatorial structure. 
This seems to suggests that the iso-depletion is fixed by the
phase-shifts, most probably via their isospin dependence. This is
corroborated at higher densities by comparing the very similar Av18 and CDBonn
predictions (left panel).

\begin{table}[t!]
  \begin{tabular}{ccc}
    \hline \hline
	Interaction & Symmetric  & Neutron \\ 
    \hline
    CDBonn      & 0.891 & 0.972 \\
    Reid93      & 0.872 & 0.962 \\
    Argonne v18 & 0.872 & 0.957 \\
    Argonne v8' & 0.863 & 0.956 \\
    Argonne v6' & 0.879 & 0.964 \\
    Argonne v4' & 0.946 & 0.971 \\
    FFG         & 0.9993 & 0.999991 \\
    \hline \hline
  \end{tabular}
\caption{Occupation of the lowest momentum state for different NN interactions in symmetric and neutron matter.}
\label{tab:deppot}
\end{table}
\begin{figure*}[t!]
  \begin{center}
    \includegraphics*[width=\linewidth]{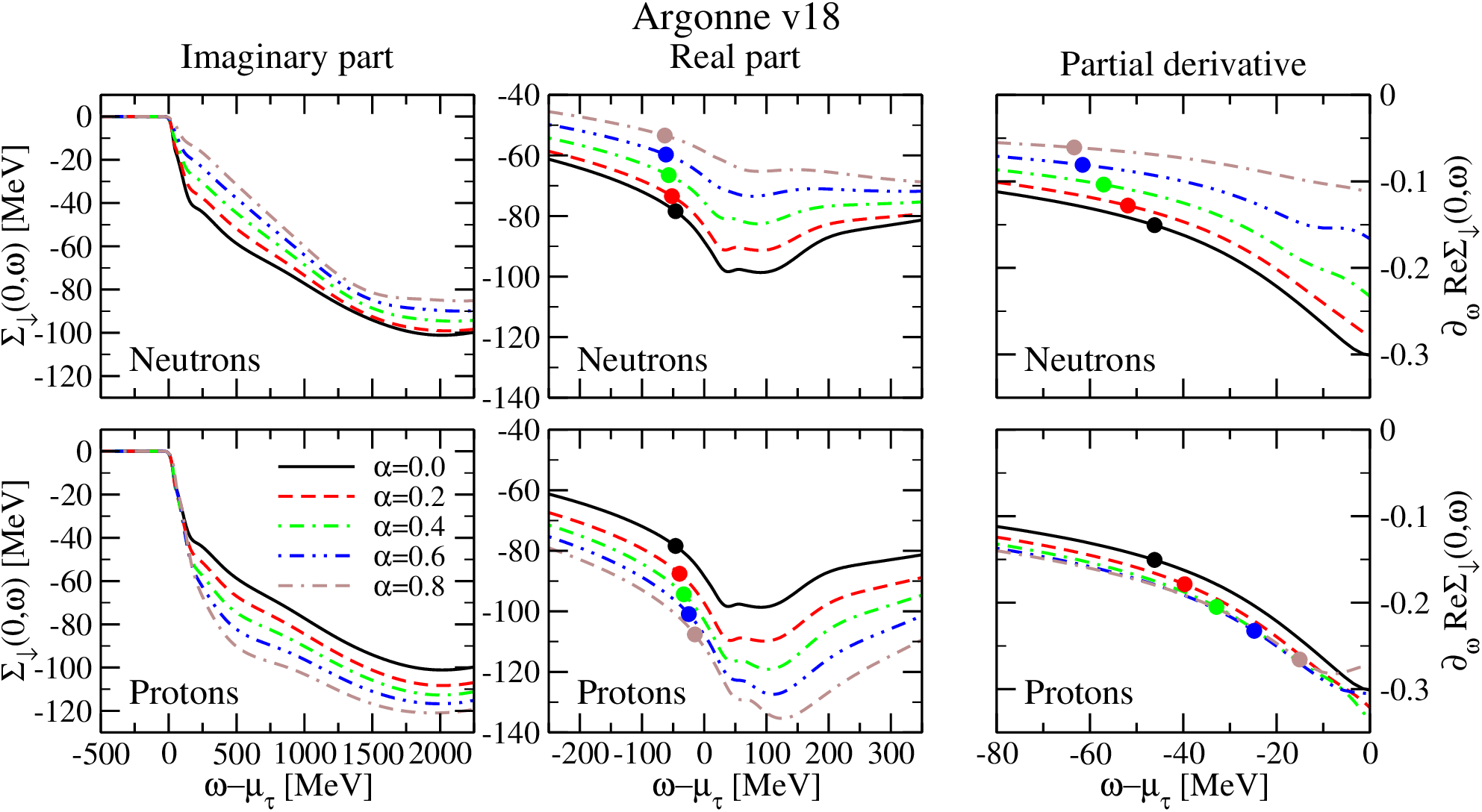}  
    \caption{(Color online) The $k=0$ imaginary (left panels) and real (central panels) parts 
of $\Sigma_\downarrow$ as a function of the energy for different asymmetries at 
$\rho=0.16$ fm$^{-3}$ and $T=5$ MeV. 
Upper (lower) panels display the results for neutrons (protons) obtained from the Av18 interaction.
The right panels display the partial derivatives of the real part with dots representing the 
location of the self-consistent quasi-particle energy.}
    \label{fig:selfdasy}
  \end{center}
\end{figure*}
In both panels, there are two results that fall below the main 
iso-depletion line. Immediately below most of the interactions, one
finds the results corresponding to the Av4' potential. This has an extremely
simplified operatorial structure, with only a spin-isospin part and no
tensor components, and is fitted to reproduce the binding energy of
the deuteron \cite{Wiringa02}. The fact that it lies significantly below the other
results shows the importance of tensor effects for isospin
asymmetric systems. It appears that the tensor force tend to increase
the difference between neutron and proton momentum distributions 
as asymmetry increases. Moreover, the comparison with Av6' and Av8'
suggests that, once the tensor components are included
in a force, the iso-depletion will remain almost the same
independently of the extra spin-orbit terms. 
Let us again stress the fact that such an agreement for different
potentials is very surprising, particularly if one considers the fact 
that the momentum distributions of neutrons and 
protons can be different for each potential. 

The FFG iso-depletion is, in all the cases explored, below the
correlated results, which shows the importance of NN interaction-induced
effects. 
Let us note that the effect of beyond mean-field correlations is essential in this case. Mean-field calculations, either with realistic or phenomenological forces, will predict results of the same order of magnitude of the FFG. 
Again, the latter provides a measure of thermal
effects. In the left panel, for instance, the
correlated iso-depletion at the largest asymmetry considered should be significantly
affected by thermal effects and therefore one should not take that
value as a NN dynamical correlation effect. For the right
panel, thermal effects are suppressed by degeneracy and the results
can be taken as purely interaction-induced, zero-temperature values.

Table~\ref{tab:deppot} summarizes the values of $n(0)$ for SNM and PNM
at $\rho=0.16$ fm$^{-3}$ and $T=5$ MeV for the different interactions
considered in the left panel of Fig.~\ref{fig:depl}. 
In all cases, $n(0)$ for PNM is larger than for SNM, confirming
the notion that, for the same density, neutron matter is less
correlated than nuclear matter. The values of the depletion for
both systems are within less than $5 \%$ of each other for the three phase-shift
equivalent potentials (and also for Av6' and Av8'), which suggests that this quantity is rather
well established at the theoretical level. Numerical uncertainties are well under
control in this case and will only affect the third significant digit
in the correlated calculations. Moreover, the difference of the occupation between neutron and nuclear matter is rather constant for all potentials. 
The conclusion is therefore warranted that the asymmetry dependence of the difference between the 
depletion for neutrons and protons is rather strongly constrained once an interaction is employed
that yields an accurate description of the NN phase shifts.
Av4' does not reproduce these and produces an
unrealistically small depletion for SNM, thus effectively decreasing
the difference between nuclear and neutron matter. 
Due to the fact that both, neutron and nuclear matter, are degenerate 
systems at this density and temperature, as indicated by the extremely
small depletions for the FFG case, these results are not
affected by thermal effects.
As a matter of fact, the results obtained with Reid93 are in very good agreement with previous calculations at 
zero temperature for the same potentials, obtained with a different procedure to implement the 
SCGF method~\cite{dewulf02}.    

The isospin asymmetry dependence of $n_\tau(0)$ can also be analyzed in terms of the time-ordered 
components of the self-energy. 
Figure~\ref{fig:selfdasy} shows the $k=0$ imaginary (left panels) and real (central panels) parts 
of $\Sigma_\downarrow$, at $\rho=0.16$ fm$^{-3}$ and $T=5$ MeV, as a function of the energy for 
different asymmetries. 
Upper (lower) panels display the results for neutrons (protons) obtained for the Av18 interaction. 
The increase of isospin asymmetry has a repulsive effect in the neutron 
$\textrm{Im} \Sigma_\downarrow$, which is shifted by almost a constant amount as $\alpha$ 
increases. 
The minimum in this function becomes shallower and the threshold at $\w \sim \mu_\tau$ is less 
pronounced.  
These repulsive effects are associated with the decreasing importance of the tensor force for 
neutrons with increasing asymmetry.
As a consequence of the diminishing threshold behavior, the inverted peak in 
$\textrm{Re} \Sigma_\downarrow$ for neutrons is smeared out at large $\alpha$'s. 
The corresponding partial derivatives (right panels) become smaller in absolute value which, together with the 
quasi-particle shift, lead to a decrease of neutron depletion as asymmetry increases. 
Let us note here that the $k=0$ quasi-particle energies for neutrons become more repulsive as 
asymmetry increases~\cite{rios06b}. 
However, the distance with respect to the (also more repulsive) chemical potential increases with 
asymmetry, providing the shift to more negative values observed in the figure. 

In the case of protons, the situation for the down component is significantly different. 
The minimum of $\textrm{Im} \Sigma_\downarrow$ becomes larger as the system becomes more neutron 
rich and the threshold is more pronounced. 
Consequently, the inverted peak in the real part becomes larger and more attractive. 
When the partial derivative is taken (right panel), similar results
are obtained for all asymmetries in 
the energy region of interest.  
This indicates that the effect of asymmetry on the proton self-energy
is mainly energy 
independent, with the real part only shifting 
by a constant at energies $\w \lesssim \mu_\tau$. 
This is in contrast to the results for neutrons (upper left panels), 
where the partial derivative changes substantially as asymmetry increases, 
\emph{i.e.} asymmetry induces energy-dependent effects for neutrons
but not for protons. 
In spite of these qualitatively different behaviors, $n_n(0)$ and $n_p(0)$ deviate from the 
symmetric value by almost the same amount as asymmetry increases (see
Fig.~\ref{fig:depasy}). 
Similar results have been obtained for other NN interactions under the same conditions.
In the view of these results, the ``universality'' found for the
iso-depletion for different interactions appears to be rather
surprising. In spite of having very different $\textrm{Im}
\Sigma_\downarrow$ components and predicting different quasi-particle
shifts, the differences in the two momentum distributions are such
that they give the same results independently of the NN force.


\section{Conclusions}
\label{sec:conclu}

The aim of the present paper is to provide a better understanding of the depletion of the nuclear Fermi sea as a function 
of density, temperature and isospin asymmetry, for different choices of realistic interactions.
The SCGF method, when implemented at the level of ladder diagrams, is capable of providing answers 
to these questions with emphasis on the role of short-range and tensor correlations that are 
induced by realistic NN interactions.
While long-range correlations have an additional influence on the depletion of the Fermi sea, it
is expected that they are more important near the Fermi energy.
For a proper comparison with finite nuclei, however, these long-range
volume effects are presumably irrelevant, since they must anyway be replaced by the 
surface dominated physics of low-lying nuclear states.
The SCGF study reported here is performed at finite temperature, on the one hand to clarify the 
role of the temperature in determining depletion effects and, on the other hand, to avoid dealing
with the technically challenging issue of pairing correlations.
Since pairing is confined to energies very near the Fermi energy, at least at densities relevant 
for nuclei, our results will not be modified when such correlations will be considered in the 
future.
We therefore report results mostly for $T=5$ and $10$ MeV, well above any pairing transition 
temperature for the systems under study~\cite{dickhoff05}.
We have studied both one-component systems (symmetric matter and pure neutron matter) as a 
function of density, as well as isospin-polarized matter as a function of asymmetry for normal
and twice normal density.

Our results indicate that temperature affects $n(k)$ in a qualitatively similar way as dynamical 
correlations do. The two types of correlations, however, induce
quantitatively different effects and can therefore be
distinguished. To separate these components, we have relied on a
comparison with the FFG results, which are only affected by
temperature effects. 
In the degenerate regime, \emph{i.e.} at high density and relatively low
temperatures, temperature effects are confined to a 
redistribution of particles around the chemical potential within an energy scale that is 
proportional to the temperature considered.
Dynamical correlations induced by the short-range NN repulsion and tensor effects, on the 
contrary, generate an almost momentum-independent depletion of the Fermi sea, and, 
complementary, lead to an occupation of high-momentum states far from the Fermi momentum.

The CDBonn and Av18 interactions have been employed to identify possible differences for the 
resulting nuclear momentum distributions.
For symmetric nuclear matter and neutron matter, the depletion for $k=0$ is strongly influenced
by temperature at low densities.
Once a density corresponding to $\rho=0.08$ fm$^{-3}$ is reached in PNM,
short-range correlations lead to slowly increasing depletion as a function of density.
This behavior is intuitive, since neutrons will be exposed more frequently to their mutual
repulsion with increasing density, thus enhancing their occupation of high-momentum states, and, 
accordingly, increasing the depletion of the Fermi sea.
For SNM, this behavior is \emph{not} observed.
The depletion saturates for the Av18 beyond normal density at a value
of about $12 \%$, 
while it continues to \emph{decrease} for CDBonn.
The larger depletion of the Av18 as compared to CDBonn further
emphasizes that the first is a slightly
harder NN interaction. This is also illustrated by the different behavior of their self-energies. 
In order to understand this puzzling density dependence, we have utilized an approximate relation between
the momentum distribution and energy derivatives of the real part of the time-ordered 
self-energy taken at the corresponding quasi-particle energy (emphasizing the importance of 
self-consistency).
For momenta below $k_F$, this result is valid when the energy dependence of the real part of the 
self-energy associated (through a dispersion integral) with the imaginary part above the chemical 
potential is essentially linear.
For momenta above $k_F$, a similar relation holds when the real part of the self-energy is 
considered that is related to the imaginary part below the chemical potential.
Near the Fermi momentum and at low densities, this approximation ceases to be valid.
However, for most situations considered in this paper, it provides an excellent approximation to
the momentum distribution calculated from the self-consistent spectral function.

In order to utilize this relation, we have generated the time-ordered contributions to the nucleon
self-energy at finite temperature. This allows us to gain access to
the appropriate energy derivatives.
Finite-temperature calculations are normally performed with retarded quantities, but the
``particle'' and ``hole'' decomposition of the self-energy provides a more suitable 
connection with the corresponding results at zero temperature.
The relevant self-energy components for determining the occupation inside the Fermi sea indeed 
shed light on the anomalous density dependence of the depletion in symmetric nuclear matter.
Mainly due to the tensor force, there is a an important threshold behavior in the imaginary part
of the self-energy above the chemical potential in symmetric matter, which is absent in
neutron matter.
For CDBonn, the threshold gives rise to an inverted peak in the corresponding real part that moves to higher energy with
increasing density, yielding an almost density independent derivative.
Since the tensor force also leads to more binding with increasing density, the location of the
derivative moves farther away from the peak with density, thereby decreasing its absolute value 
and accordingly the depletion of the Fermi sea.
For the Av18 interaction, the energy derivative decreases with density on account of its less
dramatic threshold behavior for the imaginary part, yielding similar binding from an imaginary
part that is significantly stronger than for CDBonn, but peaking at much higher energy.
Since the binding effects are similar to those for CDBonn, the two effects compensate leading
to the saturating behavior of the depletion observed in Fig.~\ref{fig:depsymneu}. 
For neutron matter, the ${}^3S_1$-${}^3D_1$ coupled partial wave is absent and no strong threshold
behavior is generated in the imaginary part for both interactions.
This leads to the more intuitive behavior that the depletion increases with the density.

For asymmetric systems, we find a significant change of depletion with isospin asymmetry. 
We identify a temperature effect for the minority protons (but not for the majority neutrons) 
at an asymmetry beyond $\alpha=0.4$ at $T=5$ MeV and normal density,
due to the corresponding low density for protons.
For twice normal density, the temperature effect does not play a role
up to an extreme asymmetry of $0.8$.
The effect was implicit (and stronger) in earlier results obtained at 
$T=10$ MeV~\cite{frick05}.
The remaining difference between the neutron and proton depletion is due to the decreased 
(increased) importance of tensor correlations for neutrons (protons). 
By considering also different operatorial clones of the Av18 interaction, like Av8', Av6', and
Av4', we can indeed unambiguously demonstrate that this difference is associated with the
tensor force. More importantly, this iso-depletion is independent of
the chosen interaction, and therefore determined solely by phase shifts at least up to twice normal density.

We close this discussion by noting that the employed interactions appear to slightly
underestimate the depletion of the experimentally determined depletion
of the deep proton 
mean-field orbits in ${}^{208}$Pb~\cite{Bat01}.
Interactions with an even stronger repulsion, like the Reid soft-core~\cite{reid68}, yield a 
slightly larger depletion of about $15 \%$~\cite{roth02,dewulf03,dickhoff04}.
Nevertheless, while there remains a few \% uncertainty as to the exact amount that is 
experimentally required, it is also clear that very different interactions, including modern 
ones, still lead to very similar predictions for the depletion of the nuclear Fermi sea.
Such depletions can be reliably calculated with the SCGF method.
When lattice QCD calculations yield unambiguous information about the short-range NN repulsion,
the remaining uncertainty about the depletion of the nuclear Fermi sea
might eventually be eliminated.

\section*{Acknowledgments}

This work was partially supported by the U.S. National Science Foundation under Grant Nos. 
PHY-0555893, PHY-0800026, and PHY-0652900 and by MICINN (Spain) under projects FIS2008-01661 
and CPAN CSD2007-00042 Programa Consolider-Ingenio 2010.

\bibliographystyle{apsrev}
\bibliography{main}

\end{document}